\documentclass[aps,prb,twocolumn,notitlepage,showpacs,floatfix,superscriptaddress]{revtex4-2}

\usepackage{bm}
\usepackage{amsmath,amssymb,graphicx}
\usepackage{bbold}
\usepackage[titletoc]{appendix}
\usepackage[colorlinks=true,citecolor=blue,urlcolor=blue,linkcolor=blue]{hyperref}
\usepackage{braket,titlesec,natbib}
\usepackage{dsfont}
\usepackage{comment}
\usepackage{float}
\usepackage{color}
\usepackage{hyperref}
\usepackage{graphicx}
\usepackage{soul}
\usepackage{ulem}
\definecolor{darkgreen}{rgb}{0.0, 0.42, 0.10}

\newcommand{\pa}[1]{\textcolor{blue}{#1}}

\newcommand{\bse}{\begin{subequations}}
\newcommand{\ese}{\end{subequations}}

\newcommand{\bs}{\boldsymbol{\sigma}}

\newcommand{\beq}{\begin{equation}}
\newcommand{\eeq}{\end{equation}}
\newcommand{\bea}{\begin{eqnarray}}
\newcommand{\eea}{\end{eqnarray}}
\newcommand{\ve}{\varepsilon}
\newcommand{\up}{\uparrow}
\newcommand{\down}{\downarrow}
\newcommand{\rigt}{\rightarrow}

\newcommand{\bk}{{\bf k}}

\newcommand{\bq}{{\bf q}}

\newcommand{\bP}{{\bf P}}

\newcommand{\bwt}{\begin{widetext}}
\newcommand{\ewt}{\end{widetext}}

\newcommand{\er}{\eqref}

\begin{document}

\title{Spin-Phonon Resonances in 
	Nearly Polar Metals with Spin-Orbit Coupling}

\author{Abhishek Kumar}
\affiliation{Department of Physics and Astronomy, Rutgers University, Piscataway, New Jersey 08854, USA}
\author{Premala Chandra}
\affiliation{Department of Physics and Astronomy, Rutgers University, Piscataway, New Jersey 08854, USA}
\author{Pavel A. Volkov}
\affiliation{Department of Physics and Astronomy, Rutgers University, Piscataway, New Jersey 08854, USA}

\begin{abstract}
In metals in the vicinity of a polar transition, interactions between electrons and soft phonon modes remain to be determined. Here we explore the consequences of spin-orbit assisted electron-phonon coupling on the collective modes of such nearly polar metals in the presense of magnetic field. We find that the soft polar phonon hybridizes with spin-flip electronic excitations of the Zeeman-split bands leading to an anticrossing. The associated energy splitting allows for an unambiguous determination of the strength of the spin-orbit mediated coupling to soft modes in polar metals by spectroscopic experiments. The approach to the polar transition is reflected by the softening of the effective $g$-factor of the hybridized spin-flip mode. Analyzing the static limit, we find that the polar order parameter can be oriented by magnetic field. This provides possibilities for new switching protocols in polar metallic materials. We demonstrate that the effects we predict can be observed with current experimental techniques and discuss promising material candidates.
\end{abstract}

\maketitle
\section{Introduction}

Though polar metals that undergo inversion-breaking transitions to phases characterized by polar space groups were predicted several decades ago \cite{Anderson:1965}, it is only relatively recently
that they have been realized experimentally \cite{Zhou_review,shi:2013,jin:2019,Fei:2018,Cao:2018,laurita:2019}
with many more predicted. \cite{Ding:2018, shirodkar:2014, Fei:2016,benedek2016}
Moreover, many anomalous properties of polar metals, particularly in the superconducting state, have been predicted \cite{kozii:2015,ruhman2017,kanasugi:2018, kanasugi:2019, kozii:2019, Maria:2020, Maria:review, gastiasoro2021theory} by invoking a spin-orbit interaction mediated coupling between polar fluctuations (arising from soft polar phonons) and electrons. However, to date there is no consensus about the magnitude of this coupling. 
A recent ab initio study provided an estimate for this type of coupling in doped strontium titanate \cite{gastiasoro2021theory}; previously its influence on superconducting state was found to be weak \cite{Ruhman:2016}.
This nearly polar metal has several unconventional properties \cite{Collignon:2019}, possibly related to quantum polar fluctuations \cite{edge2015,kumar:2021,volkov2021}.
Here we propose a mechanism to probe the strength of this spin-orbit mediated coupling in nearly polar
metals, metals in the vicinity of their polar transitions on the disordered side. We demonstrate the presence of a spin-phonon 
resonance in magnetic field; the
hybridization of the soft transverse optical phonon and the electronic spin-flip mode lead to an energy splitting that is a function of coupling strength.
Based on our results, we propose spectroscopic measurements to extract the coupling strength and we also discuss possibilities for switching in the structurally polar phase.

\begin{figure}
\centering
\includegraphics[scale=0.39]{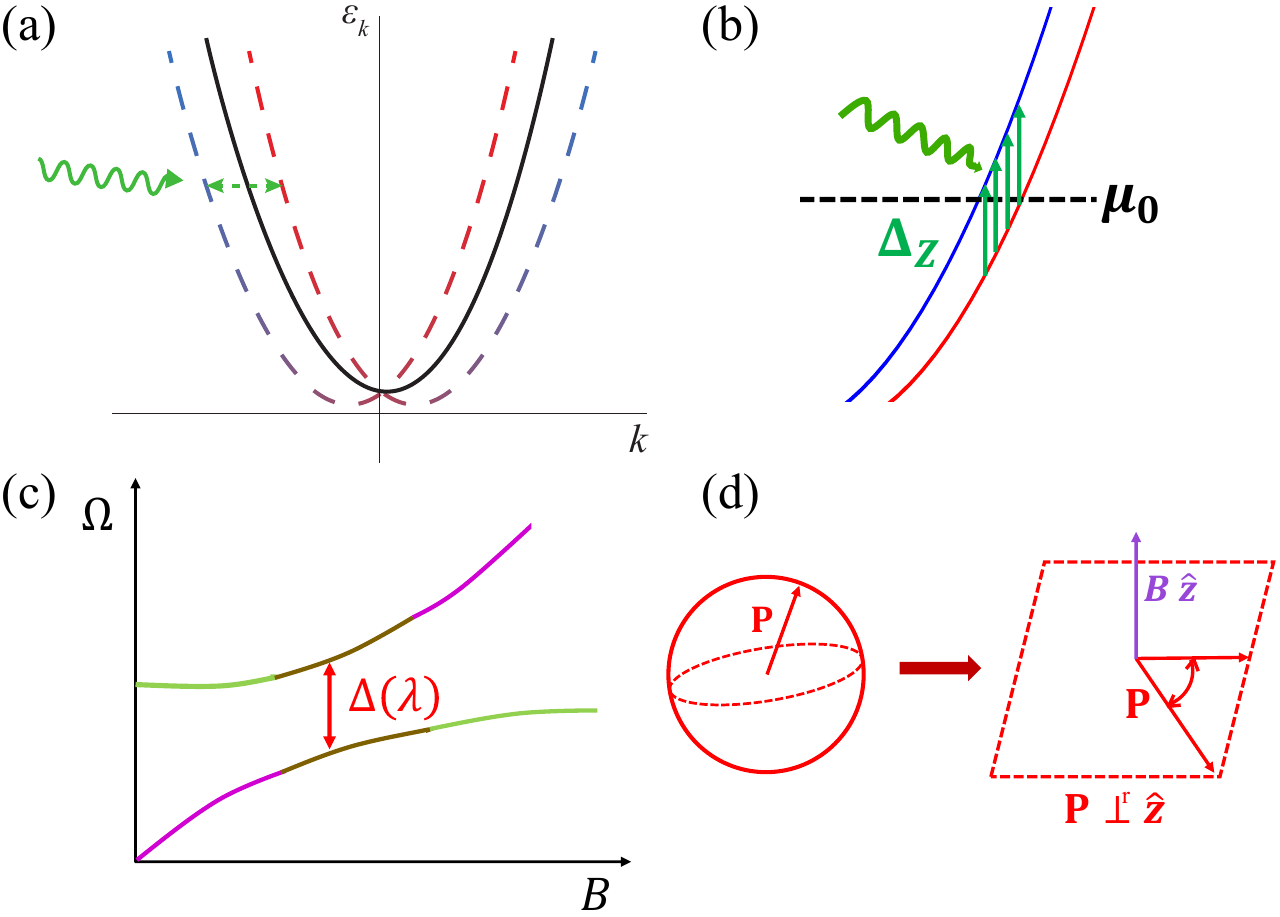}
\caption{\label{fig:intro} Schematics of (a) the virtual spin-orbit assisted electron-phonon interaction, where the green (squiggly) line is the soft TO phonon, (b) the virtual spin-orbit mediated electron-TO phonon interaction in the presence of a magnetic field near the chemical potential $\mu_0$, 
(c) the avoided crossing of the soft polar and the electronic collective modes where $\Omega$ is frequency, $B$ is magnetic field and $\Delta$ is a function of the Rashba-type electron-phonon coupling strength $\lambda$ that can be extracted experimentally, and (d) orientation of the polar order parameter by magnetic field.}
\end{figure}

In nearly polar metals, the interaction of electronic excitations and the soft phonons is an area of current exploration \cite{kozii:2015,kozii:2019,volkov2020,kumar:2021}.  Here we study
the consequences of spin-orbit assisted 
electron-phonon coupling in the vicinity of a polar critical point. The appropriate interaction Hamiltonian is 
	\beq
	\label{coupling}
	\hat{H}_\text{int} = \lambda \sum_{\bk, \bq} \sum_{s, s'} c_{\bk + \bq/2, s}^\dagger \big[ (\bk \times \hat{\bs}_{ss'}) \cdot \bP_{\bq} \big] c_{\bk - \bq/2, s'},
	\eeq
where $\lambda$ is the 
electron-phonon coupling constant, $c_{\bk, s}^\dagger (c_{\bk, s})$ is the electron creation (annihilation) operator with momentum $\bk$, spin $s = \up, \down$, $\hat{\bs}$ is the Pauli matrix for spin (or Kramers "pseudospin" quantum number) and $\bP_\bq$ describes the local polarization of the crystal at finite momentum $\bq$, which is related to the optical phonon displacement field according to the formula $\bP_\bq = n_0 (Ze) \textbf{u}_\bq$ with $n_0$ and $Ze$ as ionic density and Born effective charge, respectively. We note that \er{coupling} can be straightforwardly generalized to the case of an inversion-breaking, but non-polar phonon. In that case one should replace $\bP_\bq \to \textbf{u}_\bq$, which also results in a symmetry-allowed coupling. For the remainder of the paper we will consider mostly materials near polar instability and hence use the notation in \er{coupling}.
Here we note that the fluctuating phonon couples to the electronic spin current; since this coupling remains finite in the limit of $q \rigt 0$, it allows direct coupling to the critical mode, leading to a range of interesting phenomena emerging close to quantum criticality \cite{kozii:2015,ruhman2017, kozii:2019, Maria:2020, gastiasoro2021theory}. 
Alternatively, the interaction (\ref{coupling}) can be seen as a dynamical Rashba spin-orbit interaction, where both the polar axis direction and the coupling magnitude are dynamical fluctuating variables.
In Fig.~\ref{fig:intro}(a) we show a schematic of the dynamical Rashba effect where
the conduction bands have a dynamical shift due to interaction with the fluctuating soft phonon. 
As we show below, in the presence of a Zeeman splitting, this coupling allows for a resonant interaction between phonons and spin-interband transitions (Fig.~\ref{fig:intro}(b)). Moreover, with the application of magnetic field, one expects the spin-rotation symmetry for the conduction electrons, and consequently for the phonon polarization due to the interaction \er{coupling}, to be broken, which is otherwise preserved. Indeed, the interaction Hamiltonian \er{coupling} is invariant under simultaneous rotation of electron spin, momenta and phonon polarization. The magnetic field breaks this symmetry resulting in multiple coupled phonon and electronic modes.

The resulting coupling between spin and polar degrees of freedom raises intriguing parallels with insulating multiferroics. Here, magnetic field-induced polarization\cite{Kimura03}, 
polarization-switching\cite{Hur04}
and ferroelectricity\cite{Kim17} have been observed, offering promise for the development
of magnetically recorded ferroelectric memories.\cite{Cheong07}
Analogously in these Rashba-type polar metals, the coexistence of magnetic and polar orders, could lead to novel spintronics functionalities.
To assess such prospects of using spin-orbit coupled polar metals, the strength of $\lambda$ in  \er{coupling} is of crucial importance. However to date there have been no proposals to directly measure the magnitude of this spin-orbit assisted electron-phonon coupling constant.

In this paper, we study the collective modes of a nearly polar metal in the presence of a magnetic field. 
We demonstrate that the spin-orbit mediated interaction \eqref{coupling} leads to directly measurable spectroscopic signatures: avoided crossing (anticrossing) of the spin-flip modes of conduction electrons and the coupled phonon-plasmon modes (see Fig.~\ref{fig:intro}(c)). 
The magnitude of the coupling can be extracted from the splitting emerging at the anticrossing point. Furthermore the
collective electronic spin-flip modes softens as the polar
critical point is approached.
In the ordered phase, we show that the spin-orbit assisted electron-phonon coupling in \eqref{coupling} makes it possible to orient the polar order with magnetic field, 
aligning it in the plane perpendicular to the field (see Fig.~\ref{fig:intro}(d)). Finally we discuss specific experiments that can extract the spin-orbit assisted coupling between the polar mode and the electrons, and possible spintronics functionalities.

The rest of the paper is organized as follows. 
In Sec.~\ref{model} we describe our model and our general strategy for studying collective modes in magnetic field in nearly polar metals, which includes the discussion of effects of long-range Coulomb repulsion on electrons and polar phonons.
The hybridization of phonons, plasmons and electronic spin-flip modes at finite energies is discussed in Secs.~\ref{SE_q=0} and \ref{hybrid}.
In Sec.~\ref{finite_q} we explore the effects of magnetic field on the order parameter and the critical low-energy properties near the polar transition. 
In Sec.~\ref{discuss} we discuss the effects of temperature, disorder and orbital quantization on our results (\ref{approx}), results for 2D systems (\ref{2D}), and the experimental proposals for extracting the strength of spin-orbit assisted electron-phonon coupling and orienting the polar order with magnetic field (\ref{exp}). Our results are summarized in Sec.~\ref{conclu}.

\section{Model and general relations}
\label{model}
In this section we describe the model we use to study the collective modes of a spin-orbit coupled nearly polar metal in the presence of magnetic field. In Sec. \ref{eei} we discuss the effects of Coulomb interactions, in particular the distinction between LO and TO modes that it introduces, and in Sec. \ref{el-ph} we present general relations used to evaluate the collective modes in the presence of coupling \er{coupling}.

To facilitate the analytical progress, we consider a single parabolic band model for conduction electrons. For most derivations, we will take $T=0$  and ignore the orbital effect of the magnetic field, focusing only on the Zeeman effect for conduction electrons (see Sec.~\ref{approx} for the effects of finite temperatures and orbital quantization which do not change the qualitative results). We further assume (without the loss of generality due to spherical symmetry of the model) the magnetic field to be applied along $z$-direction.
The electronic Hamiltonian is written as
\beq
\hat H = \hat{H}_0 + \hat{H}_\text{int},
\eeq
where $\hat{H}_\text{int}$ is given by \eqref{coupling} and the single particle Hamiltonian $\hat{H}_0$ is given by
\beq
\label{single}
\hat{H}_0 = \sum_{\bk, s} \ve_\bk c_{\bk, s}^\dagger c_{\bk, s}  + \frac{1}{2} g \mu_B \sum_{\bk, s, s'} \textbf{B} \cdot \hat{\bs}_{ss'} c_{\bk, s}^\dagger c_{\bk, s'},
\eeq
where $\ve_\bk = k^2/2m_b$ is the free particle dispersion, $\textbf{B} = B \hat{z}$ is the magnetic field, $g$ is the Lande-$g$ factor, $\mu_B$ is the Bohr magneton and $m_b$ is the band mass. Additionally, we take the effects of Coulomb interaction into account, which is discussed in detail in Sec.~\ref{eei}.

We now discuss the bare phonon part of the system. The order parameter field $\bP_{\bq}$ can be represented in terms of phonon creation and annihilation operators:\cite{kittelsolids}
\beq
\label{pol}
\bP_{\bq} = \frac{1}{\sqrt{V}} \bigg[ \frac{\omega_\bq (\ve(\bq) - \ve_\infty)}{4\pi} \bigg]^{1/2} \big[ b_\bq e^{-i \omega_\bq t} + b_\bq^\dagger e^{i \omega_\bq t} \big] \bf{e}_\bq,
\eeq
where $\omega_\bq$ is the phonon dispersion, $\ve(\bq)$ and $\ve_\infty$ correspond to static and high frequency dielectric constants in the absence of free carriers, $\bf{e}_\bq$ is unit polarization vector and $V$ is volume of the crystal.

To study the effects of interaction, we require the form of the phonon propagator. The bare phonon Green's function can be obtained by calculating the correlation of order parameter fields $\bP_\bq$ \er{pol} in the Matsubara formalism:
\beq
\label{TO}
\mathcal{D}_{\alpha\beta}^0 (\bq, \Omega_m) = \frac{1}{V} \frac{\omega_\bq (\ve(\bq) - \ve_\infty)}{4\pi} \frac{-2\omega_\bq}{\Omega_m^2 + \omega_\bq^2} e_\alpha(\bq) e_\beta(\bq).
\eeq
Using $(\ve(\bq) - \ve_\infty) = \ve_\infty \omega_{pi}^2/\omega_\bq^2$,\cite{mahan:book} where $\omega_{pi} = \sqrt{4\pi (Ze)^2 n_0/(\ve_\infty M)}$ is the ionic plasma frequency with $n_0$ as the density of ion pairs, $Ze$ as the ionic charge and $M$ as the reduced ionic mass, the Eq.~\er{TO} can be written as
\beq
\label{green}
\mathcal{D}_{\alpha\beta}^0 (\bq, \Omega_m) = - \frac{1}{V} \frac{\Omega_0^2}{2\pi} \frac{1}{\Omega_m^2 + \omega_\bq^2} e_\alpha(\bq) e_\beta(\bq),
\eeq
with $\Omega_0 \equiv \sqrt{\ve_\infty} \omega_{pi}$. For brevity, we will write $\mathcal{D}_{\alpha\beta}^0(\bq, \Omega_m)$ as $\mathcal{D}_{\alpha\beta}^0$ in what follows. We note that the phonon dispersion ($\omega_\bq$) appearing in the Green's function \er{green} does not distinguish between LO and TO modes. For the same reason, the unit polarization vector $\textbf{e}_\bq$ obeys $e_\alpha(\bq)e_\beta(\bq) = \delta_{\alpha\beta}$. The distinction between LO and TO modes amounts to splitting the total polarization into that for longitudinal and transverse ones, with the dispersion $\omega_\bq$ also being distinguished as that for LO and TO modes, correspondingly. Later in this section we will discuss the LO-TO distinction originating from the long range Coulomb interaction.

\subsection{Effects of Coulomb interactions near the polar critical point}
\label{eei}
We now discuss properties of the Coulomb interaction near a polar critical point (PCP). The Coulomb repulsion experienced between two electrons in a doped polar semiconductor is screened by both the polar phonons and electrons and is given by
\beq
\label{coul1}
V(\bq, \Omega_m) = \frac{4\pi e^2}{\ve(\bq, \Omega_m) q^2},
\eeq
where
\beq
\label{diel}
\ve(\bq, \Omega_m, T) = \ve_\infty + \frac{\ve_\infty \omega_{pi}^2}{\Omega_m^2 + \omega_\text{TO}^2(\bq, T)} + \frac{4\pi e^2}{q^2} \mathcal{P}(\bq, \Omega_m)
\eeq
is the dielectric permittivity given by the sum of its lattice and free electron parts.\cite{mahan:book} Here, as mentioned above, $\omega_{pi}$ is the ionic plasma frequency, $\omega_\text{TO}^2 (\bq, T) \equiv \omega_\bq^2 \approx \omega_\text{TO}^2(0, T) + c_T^2 q^2$ is the TO phonon dispersion near the $\Gamma$-point of the Brillouin zone, with $\omega_\text{TO}(0, T)$ being the temperature $(T)$-dependent TO mode frequency and $c_T$ being the speed of TO mode, and
\beq
\label{susc}
\mathcal{P}(\bq, \Omega_m) = \nu_F \Big( 1 - \frac{\Omega_m}{v_F q} \text{arctan} \frac{v_F q}{\Omega_m} \Big)
\eeq
is the polarizability of charge carriers for $q \ll k_F$ in the degenerate regime considered within random phase approximation (RPA, justified below), with  $\nu_F$ as the electronic density of states.

In the absence of conduction electrons, the energy of the zone-center TO mode $\omega_\text{TO}^2(0, T) \propto 1/\ve_0(T)$, where $\ve_0(T)$ is the static dielectric permittivity which diverges at the PCP. This suggests $\omega_\text{TO}(0, T \rigt T_c) \to 0$ near the PCP. $T_c$ can be manipulated by a number of means such as stress\cite{Burke1971, Uwe1976}, isotope exchange,\cite{itoh_1999} or isovalent doping\cite{Mitsui1961, Rischau2017, Engelmayer2019b} and reached at the quantum limit ($T_c \to 0$) to be called as the polar (quantum) critical point.

\subsubsection{Weakness of electron-electron interaction near the polar transition}
\label{eei_weak}

Near the PCP, the Coulomb repulsion between electrons is weak due to diverging contribution of the dielectric permittivity to the soft phonons.\cite{volkov2021} In particular, the usual parameter describing the importance of Coulomb correlations is $r_s^*\sim 1/(k_F a_B \varepsilon_0(\bq))$ in the presence of phonon screening, where $q$ is of the order of the inverse average interelectron distance, i.e. $k_F$. For metals with large $k_F$, the lower bound for $\varepsilon_0(k_F)$ is $(\omega_\text{LO}^\text{min})^2/(\omega_\text{TO}^\text{max})^2$, where $\omega_\text{TO}^\text{max}$ is the maximal energy of the TO phonon in the Brillouin zone and $(\omega_\text{LO}^\text{min})^2 = (\omega_\text{TO}^\text{min})^2 + \omega_{pi}^2$. For sufficiently large Born charge of the polar mode, $\omega_{pi}\gg\omega_\text{TO}$, and the dispersion of the LO mode (as well as the effects of anisotropy at finite momenta) can be neglected. This results in $\varepsilon_0(k_F)\sim\omega_{pi}^2/(\omega_\text{TO}^\text{max})^2\gg1$, which implies $r_s^*\ll1$. For lower electron concentrations, the situation is even more straightforward. Sufficiently close to the PCP, $\varepsilon_0(k_F)\sim \omega_{pi}^2/c_T^2 k_F^2$, suggesting $r_s^*$ to be vanishingly small as $k_F\to0$. Consequently, the effects of the Coulomb repulsion on the electrons (e.g., self-energy) can be neglected for a nearly polar metal with a sufficiently large phonon Born effective charge at any carrier concentration.

\subsubsection{Distinction between TO and LO mode frequencies in the presence of free charge carriers}
\label{distinction}

In insulating ferroelectrics, long-range dipolar interactions lead to a splitting between LO and TO modes. To assess this effect in a polar metal, we consider the Coulomb interaction \er{coul1} and substitute Eq.~\er{diel} for the dielectric permittivity into Eq.~\er{coul1} to obtain the screened Coulomb interaction between two conduction electrons as
\beq
\label{coul}
V(\bq, \Omega_m) = \frac{4\pi e^2}{\ve_\infty q^2 + 4\pi e^2 \mathcal{P}(\bq, \Omega_m)} \frac{\Omega_m^2 + \omega_\text{TO}^2(\bq)}{\Omega_m^2 + \omega_\text{LO}^2(\bq, \Omega_m)},
\eeq
where
\beq
\label{distin}
\omega_\text{LO}^2(\bq, \Omega_m) = \omega_\text{TO}^2(\bq) + \frac{q^2 \omega_{pi}^2}{q^2 + \frac{4\pi e^2}{\ve_\infty} \mathcal{P}(\bq, \Omega_m)},
\eeq
with $\omega_\text{TO}(\bq)$ as the TO phonon dispersion near the $\Gamma$-point of the Brillouin zone. The LO mode frequency, $\omega_\text{LO}^2(\bq)$, is the pole of Eq.~\er{coul} which can be found by solving $\omega_\text{LO}^2(\bq, -i \Omega) - \Omega^2 = 0$ (after analytic continuation) for $\Omega$, where $\omega_\text{LO}^2(\bq, -i \Omega)$ is given by Eq.~\er{distin}. The solution is discussed in detail in Sec. \ref{hybrid_Z=0}; here we discuss the qualitative aspects of the distinction between LO and TO phonons.



One can split the phonon propagator into contributions of TO and LO modes using the identity $e_\mu(\bq) e_\nu(\bq) = \delta_{\mu\nu}$. For transverse and longitudinal polarizations, this identity can be split into
\beq
e_\mu(\bq) e_\nu(\bq) = \bigg[ \delta_{\mu\nu} - \frac{q_\mu q_\nu}{q^2} \bigg] + \bigg[ \frac{q_\mu q_\nu}{q^2} \bigg],
\eeq
where the first term corresponds to the polarization for TO modes while the second term for LO modes.
Accordingly, the phonon dispersion in Eq.~\er{green} is replaced by that of TO and LO modes, respectively.
Upon considering this distinction, the phonon Green's function is written as
\beq
\label{green1}
\begin{split}
-\frac{\Omega_0^2}{2\pi V} \big[ \mathcal{D}_{\alpha\beta}^0 \big]^{-1} = \big[ \Omega_m^2 &+ \omega_\text{TO}^2(\bq) \big] \delta_{\alpha\beta} \\
&+ \big[ \omega_\text{LO}^2(\bq, \Omega_m) - \omega_\text{TO}^2(\bq) \big] \frac{q_\alpha q_\beta}{q^2}.
\end{split}
\eeq
We now show that the LO-TO distinction (second term on the RHS of Eq.~\er{green1}) depends sensitively on the concentration of conduction electrons. In the absence of screening by conduction electrons one recovers the usual expression for LO phonon frequency, different from that of the TO one:
\beq
\label{LO}
\omega_\text{LO}^2(\bq) = \omega_\text{TO}^2(\bq) + \omega_{pi}^2.
\eeq
Let us now consider the effect of conduction electrons. For $\Omega \gg v_Fq$, where $\Omega$ is the retarded frequency, we can write Eq.~\er{susc} as $\mathcal{P}(\bq, \Omega_m) \approx (\nu_F/3) (v_Fq/\Omega_m)^2$, which subsequently gives Eq.~\er{distin} to be
\beq
\label{distin1}
\omega_\text{LO}^2(\bq, \Omega_m) \approx \omega_\text{TO}^2(\bq) + \frac{\omega_{pi}^2}{1 + \omega_{p\infty}^2/\Omega_m^2},
\eeq
where 
\beq
\omega_{p\infty} = \sqrt{\frac{4\pi n e^2}{\ve_\infty m_b}}
\eeq
is the plasma frequency with $n$ as the carrier concentration and $m_b$ as the electron effective mass. We note that it is the high frequency limit of the dielectric function, i.e., $\ve_\infty$, that enters into plasma frequency here, not $\ve_0$, the static one.

Eq.~\er{distin1} suggests that the LO-TO splitting would be negligible when $\omega_{pi}$ is small. The relevant scale for comparison here is $\omega_{p\infty}$ which can be tuned by carrier density. Qualitatively, if we assume that if the carrier density is high such that $\Omega \sim \omega_\text{LO} \ll \omega_{p\infty}$, then Eq.~\er{distin1} can be written as
\beq
\label{wt<wp}
\omega_\text{LO}^2(\bq) \sim \omega_\text{TO}^2(\bq) \bigg( 1 + \frac{\omega_{pi}^2}{\omega_{p\infty}^2} \bigg).
\eeq
Furthermore, if $\omega_{p\infty} \gg \omega_{pi}$, then we can ignore the second term of Eq.~\er{wt<wp} which means $\omega_\text{LO}(\bq) \equiv \omega_\text{TO}(\bq)$, i.e., no distinction between LO and TO modes. So the poles of Green's function \er{green1} suggest a triple degenerate phonon mode and a large energy plasmon, $\omega_{p\infty}$.



Finally, if we assume that $\Omega \ll v_Fq$ (or effectively the static limit, to be considered in Sec.~\ref{finite_q}), then $\mathcal{P}(\bq, \Omega_m=0) \approx \nu_F$, which simply implies static screening by conduction electrons. This gives the difference between phonon frequencies \eqref{distin} to be
\beq
\label{diff2}
\omega_\text{LO}^2(\bq, \Omega \ll v_F q) - \omega_\text{TO}^2(\bq) \approx \frac{q^2 \omega_{pi}^2}{q^2 + \kappa^2},
\eeq 
where we recall that $\kappa = \sqrt{4\pi e^2\nu_F/\ve_\infty}$ is the inverse screening length. The Eq.~\er{diff2} suggests that the LO-TO splitting in this regime cannot be ignored at finite $q$.

\subsection{Electron-phonon interaction and collective modes}
\label{el-ph}
Having justified that the electron-electron interaction is weak near the PCP in Sec.~\ref{eei_weak}, we consider electron-phonon interaction of Rashba SOC type \er{coupling} where the polar order fluctuation field $\bP_\bq$ couples to electron density at linear order.\cite{kanasugi:2018, kanasugi:2019, kozii:2019, Maria:2020, Maria:review, Fu:2015, kozii:2015}

To study the eigenmodes emerging in the interacting nearly-polar metal, we study phonon self-energy due to the electron-phonon coupling \er{coupling}. Assuming weak coupling, we limit ourselves to the lowest-order (in $\lambda$) self-energy. As we approach to the polar transition from the disordered side, we assume the characteristic TO mode frequency $\omega_\text{TO}(0, T)$ to be finite and large enough, such that the effects of critical fluctuations can be neglected \pa{(see Appendix \ref{app:pert})}.
The analogue of the Dyson equation for the full phonon propagator is
\beq
\label{bareTO}
\hat{\mathcal{D}}(\bq, \Omega_m) = \big[ [\hat{\mathcal{D}}^0(\bq, \Omega_m)]^{-1} - \hat{\Pi}(\bq, \Omega_m) \big]^{-1},
\eeq
where $\hat{\Pi}(\bq, \Omega_m)$ is the phonon self-energy and $\hat{\mathcal{D}}^0(\bq, \Omega_m)$ is the bare phonon propagator given in Eq.~\er{green1}. The collective modes are obtained by solving  $\text{Det}[\hat{\mathcal{D}}^{-1}(\bq, \Omega )] = 0$ for $\Omega$, where analytic continuation to real frequencies $i\Omega_m\to\Omega+ i0^+$ is assumed.

The lowest-order self-energy correction to the bare phonon propagator with the interaction vertex as $(\bk \times \bs)_i$ from Eq.~\er{coupling} can be explicitly written as
\beq
\begin{split}
\label{TO_se}
\Pi_{\alpha\beta} &(\bq, \Omega_m) = \lambda^2 T \sum_{\omega_n} \sum_\bk \text{Tr} \Big[ (\bk \times \hat \bs)_\alpha \times \\
\times &\hat G \Big( \omega_n - \frac{\Omega_m}{2}, \bk - \frac{\bq}{2} \Big) (\bk \times \hat \bs)_\beta \hat G \Big( \omega_n + \frac{\Omega_m}{2}, \bk+ \frac{\bq}{2} \Big) \Big],
\end{split}
\eeq
where $\alpha, \beta ~ \epsilon ~ (1...3)$ and 
\beq
\label{el_gr}
\hat G(\epsilon_n, \bk) = \frac{1}{2} \sum_s [\hat \sigma_0 + s \hat \sigma_z] \frac{1}{i\epsilon_n - \xi_k^s},
\eeq
is a single particle Green's function for electrons with $\xi_k^s = \ve_\bk + s \Delta_Z/2 - \mu(\Delta_Z)$ and $s = \pm 1$. Here, $\Delta_Z = g\mu_B B$ is the Zeeman energy and $\mu(\Delta_Z)$ is a chemical potential which itself is a function of magnetic field; the technical details of obtaining $\mu(\Delta_Z)$ is deferred to Appendix~\ref{chem_B}. The appearance of $\hat{\sigma}_z$ in the Green's function is due to magnetic fields applied along $z$-direction. Note that the magnetic field affects directly only the electronic dispersion, splitting it into spin-up and spin-down subbands. Consequently, the magnetic field effects on the polar order in our model are directly related to the coupling \er{coupling}.

The form of self-energy, after the trace is taken and the frequency summation, can be written as
\beq
\label{se1}
\begin{split}
\Pi_{\alpha\beta} (\bq, \Omega_m) =& \frac{\lambda^2}{2} \sum_{s \bar{s}} \sum_\bk k^2 f_{\alpha\beta}^{s \bar{s}}(\theta, \phi) \times \\
& \times \frac{n_F(\xi_{\bk-\bq/2}^{\bar{s}}) - n_F(\xi_{\bk+\bq/2}^s)}{i\Omega_m - \ve_{\bk+\frac{\bq}{2}} + \ve_{\bk-\frac{\bq}{2}} - (s-\bar{s})\frac{\Delta_Z}{2}}
\end{split}
\eeq
where $n_F(\xi_\bk^r)$ is the Fermi function, $\sum_\bk \rigt V \int d^3k/(2\pi)^3$ in the continuum limit, and
\beq
\label{coh_fac}
\begin{split}
f_{xx}^{s \bar{s}} (\theta, \phi) &= (1+s\bar{s}) \sin^2\theta \sin^2\phi + (1-s\bar{s}) \cos^2\theta, \\
f_{yy}^{s \bar{s}} (\theta, \phi) &= (1+s\bar{s}) \sin^2\theta \cos^2\phi + (1-s\bar{s}) \cos^2\theta, \\
f_{zz}^{s \bar{s}} (\theta, \phi) &= (1-s\bar{s}) \sin^2\theta, \\
f_{xy}^{s \bar{s}} (\theta, \phi) &= -(1+s\bar{s}) \sin^2\theta \frac{\sin2\phi}{2} + i (s-\bar{s}) \cos^2\theta, \\
f_{yx}^{s \bar{s}} (\theta, \phi) &= f_{xy}^{s \bar{s}} (\theta, \phi) [i \rigt -i], \\
f_{xz}^{s \bar{s}} (\theta, \phi) &= -(1-s\bar{s}) \frac{\sin2\theta}{2} \cos\phi - i (s-\bar{s}) \frac{\sin2\theta}{2} \sin\phi, \\
f_{zx}^{s \bar{s}} (\theta, \phi) &= f_{xz}^{s \bar{s}} (\theta, \phi) [i \rigt -i], \\
f_{yz}^{s \bar{s}} (\theta, \phi) &= -(1-s\bar{s}) \frac{\sin2\theta}{2} \sin\phi + i (s-\bar{s}) \frac{\sin2\theta}{2} \cos\phi, \\
f_{zy}^{s \bar{s}} (\theta, \phi) &= f_{yz}^{s \bar{s}} (\theta, \phi) [i \rigt -i],
\end{split}
\eeq
where $\theta$ and $\phi$ are polar and azimuthal angles of $\bk$, respectively. While writing Eq.~\er{se1}, we assumed the system to be partially polarized when both the magnetically split spin-up and spin-down subbands are (partially) occupied. For future purposes, we refer this scenario to be the ``two-band" case. The fully polarized system, when only the lowest subband (spin-down) is occupied, is referred to as ``one-band" case. Unless specifically mentioned, all our discussion is for the two-band case.

\section{Spin-phonon-plasmon resonances at $q=0$}
\label{zero_q}

In this section we present the results on the spectrum of collective modes at low momenta and finite frequencies (in the limit $\Omega \gg v_F q$) as a function of magnetic field. Without the interaction \er{coupling}, the system hosts a doubly-degenerate critical TO mode, two coupled plasmon-LO modes, and a precession-like spin-flip resonance of conduction electrons with energy $\Delta_Z = g\mu_B B$. The interaction leads to hybridization between these modes, resulting in the anticrossing at fields where their frequencies coincide. The latter can be used to extract the value of the coupling constant $\lambda$ from spectroscopic experiments.

Finally, we demonstrate that on approach to the transition, the effective $g$-factor of the spin-flip mode softens. We further examine the properties of the polar transition for the interacting system in Sec. \ref{finite_q}.

\subsection{Phonon self-energy at $q=0$}
\label{SE_q=0}
As a first step towards calculating the collective modes, we calculate the leading order self-energy correction at $q=0$. 
The components of the self-energy can be obtained by substituting $q=0$ in Eq.~\eqref{se1} and performing the $\bk$-integral. The explicit calculation at $T=0$ gives
\beq
\label{Pi}
\hat{\Pi}(\Omega_m) = - \frac{\lambda^2 m_b^{5/2} V}{15 \pi^2} \frac{L\big[ \mu(\Delta_Z), \Delta_Z \big]}{\Omega_m^2 + \Delta_Z^2} \hat{\tilde{\Pi}},
\eeq
where
\beq
\label{Pi_til}
\hat{\tilde{\Pi}} =
\begin{pmatrix}
\Delta_Z & -\Omega_m & 0 \\
\Omega_m & \Delta_Z & 0 \\
0 & 0 & 2\Delta_Z
\end{pmatrix}
\eeq
and 
\beq
\begin{gathered}
\label{L}
L\big[ \mu(\Delta_Z), \Delta_Z \big] = 
\\
=
\Big[ \big( 2\mu(\Delta_Z) + \Delta_Z \big)^{5/2} - \big( 2\mu(\Delta_Z) - \Delta_Z \big)^{5/2} \Big] \equiv L_Z.
\end{gathered}
\eeq
The form of $L\big[ \mu(\Delta_Z), \Delta_Z \big]$ given here is for the two-band case; for the one-band case, the second term of Eq.~\er{L} has to be omitted. From now on, for brevity, we will refer $L\big[ \mu(\Delta_Z), \Delta_Z \big]$ as $L_Z$. 

We now discuss the $q=0$ results \er{Pi} using the symmetries of the expression for the phonon self-energy \er{TO_se}.
The phonon self-energy \eqref{TO_se} can be identified as a linear combination of correlators between different components of spin-current along different directions, which are related to the response functions as per Kubo formula. Indeed, the spin-current operator is defined as $\hat{j}_i^{\sigma_a} = k_i \hat{\sigma}_a$, which is understood as the $i^\text{th}$-component of spin-current along $a$-direction. Upon explicitly writing one of the components of the self-energy, one can easily identify it to be a combination of several spin current-current correlation functions. For instance, $\Pi_{xx}$ consists of $j_y^{\sigma_z} - j_y^{\sigma_z}$, $j_z^{\sigma_y} - j_z^{\sigma_y}$, $j_y^{\sigma_z} - j_z^{\sigma_y}$ and $j_z^{\sigma_y} - j_y^{\sigma_z}$ correlations.

In the absence of magnetic field, the phonon self-energy \er{TO_se}, interpreted as a combination of spin current-current correlators, vanishes at $q=0$. Indeed, in the absence of magnetic field, the spin-current is a conserved quantity at $q=0$. So any combination of the spin current-spin current response functions must vanish in this limit, explaining why the phonon self-energy is zero in this case.

Once the magnetic field is applied along $z$-direction, only the spin-current along $z$, i.e., $j_i^{\sigma_z}$, commutes with terms having spin-current along $z$-direction in the interaction Hamiltonian \er{coupling}. In the self-energy, these kind of terms either trace out identically or give intra-band contributions, $(1+s\bar{s})$, which has zero spectral weight at $q=0$ in agreement with $j_i^{\sigma_z}$ conservation. The terms with spin-current along $x$- and $y$-directions in Eq.~\er{coupling}, however, do not commute with $j_i^{\sigma_z}$ and give finite contribution to the self-energy. This leaves only the $j_z^{\sigma_y} - j_z^{\sigma_y}$ correlation contributing to $\Pi_{xx}$, corresponding to the second term of $f_{xx}^{s \bar{s}}(\theta, \phi)$ in Eq.~\er{coh_fac}.
Due to rotational symmetry in the $xy$-plane, $\Pi_{yy}$ is identical to $\Pi_{xx}$. However, physically it's $j_z^{\sigma_x} - j_z^{\sigma_x}$ correlation that contributes to $\Pi_{yy}$.

Finally, $\Pi_{zz}$ consists of $j_x^{\sigma_y} - j_x^{\sigma_y}$, $j_y^{\sigma_x} - j_y^{\sigma_x}$, $j_x^{\sigma_y} - j_y^{\sigma_x}$ and $j_y^{\sigma_x} - j_x^{\sigma_y}$ correlations. Out of these four, only direct correlations, $j_x^{\sigma_y} - j_x^{\sigma_y}$ and $j_y^{\sigma_x} - j_y^{\sigma_x}$, are nonzero: cross correlations, $j_x^{\sigma_y} - j_y^{\sigma_x}$ and $j_y^{\sigma_x} - j_x^{\sigma_y}$, identically cancel each other according to Pauli spin anti-commutation principle.
 Parenthetically, we note that the cross correlations in $\Pi_{zz}$ do not vanish due to rotational symmetry in the $xy$-plane as the coherence factor $f_{zz}^{s \bar{s}}(\theta, \phi)$ in Eq.~\eqref{coh_fac} does not include any parity breaking term (or odd in $\phi$ terms). Therefore, $\Pi_{zz}$ has contributions from two non-zero direct correlations, while $\Pi_{xx}$ and $\Pi_{yy}$ have only one. This indicates that the weight of $\Pi_{zz}$ is twice that of $\Pi_{xx}$ or $\Pi_{yy}$, in agreement with explicit results in Eqs.~\er{Pi} and \er{Pi_til}.

Along the same lines, one can argue about off-diagonal components of the self-energy as well. The $\Pi_{xy}$ component is a combination of $j_y^{\sigma_z} - j_x^{\sigma_z}$, $j_z^{\sigma_y} - j_z^{\sigma_x}$, $j_y^{\sigma_z} - j_z^{\sigma_x}$ and $j_z^{\sigma_y} - j_x^{\sigma_z}$ correlations. Out of these only $j_z^{\sigma_y} - j_z^{\sigma_x}$ correlation yields a nonzero contribution. 
The other off-diagonal components, $\Pi_{xz}$ and $\Pi_{yz}$, vanish completely due to trace and the rotational symmetry in the $xy$-plane, as evident from the forms of $f_{xz}^{s \bar{s}}(\theta, \phi)$ and $f_{yz}^{s \bar{s}}(\theta, \phi)$ in Eq.~\eqref{coh_fac}.

\subsection{Mode hybridization and approach to the transition}
\label{hybrid}
The next step is to calculate eigenmodes of the system. We will first discuss the $\Delta_Z=0$ case. In this case we reproduce the known results for coupled phonon and electronic plasmon modes. Next, we will consider finite $\Delta_Z$ which brings spin-flip transition modes into consideration. The most important aspect of the coupling of spin transitions to phonons and plasmons is the appearance of anticrossing as a function of magnetic field, which allows to extract the electron-phonon coupling strength ($\lambda$) from spectroscopic experiments. More detailed discussion of possible experimental proposals to extract the coupling strength is given in Sec.~\ref{exp}.

\subsubsection{Phonon-plasmon coupling at $\Delta_Z=0$}
\label{hybrid_Z=0}
\begin{figure}
\centering
\includegraphics[scale=0.37]{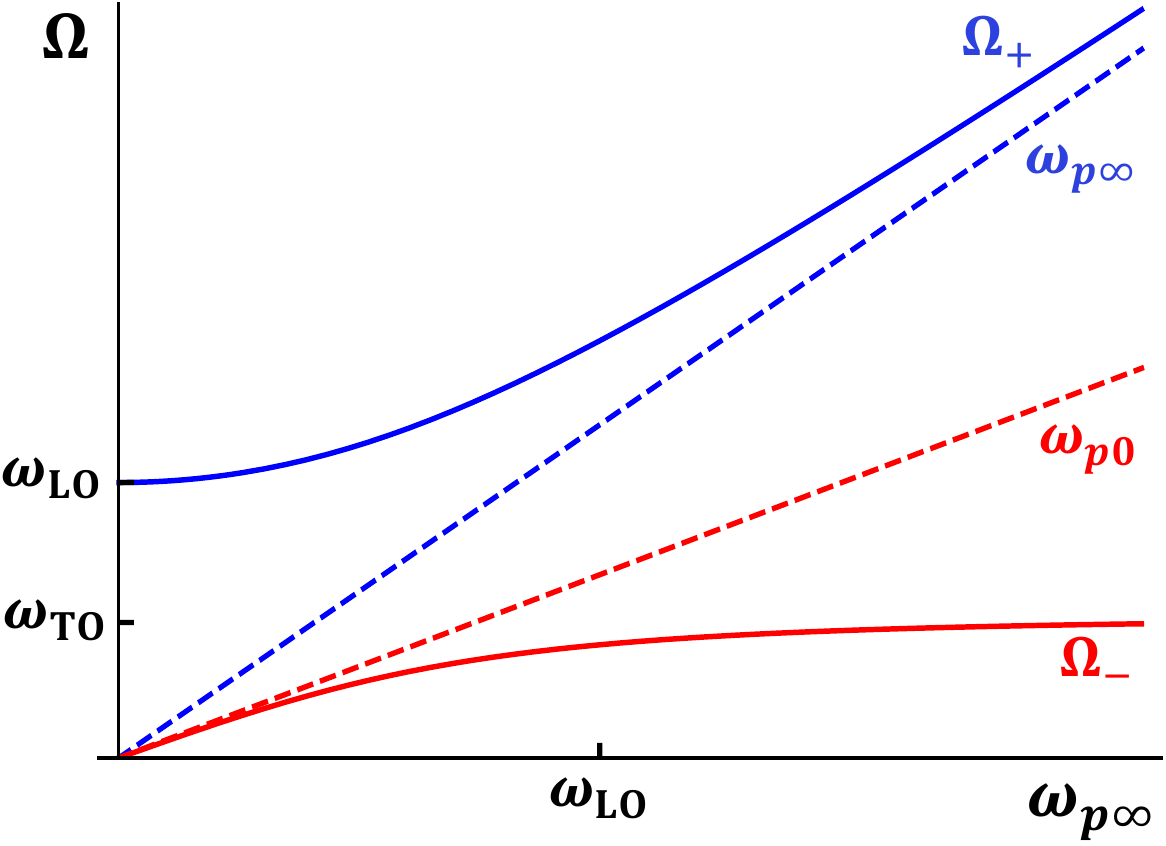}
\caption{\label{hyb_Z=0} The frequencies of the hybridized phonon-plasmon modes $\Omega_+$ and $\Omega_-$ as a function of bare plasmon frequency $(\omega_{p\infty})$, which grows with electron density. 
At high carrier densities $(\omega_{p\infty} \gg \omega_\text{LO})$, the electron gas screens the Coulomb interaction which in turn makes the LO-TO splitting disappear. The phonon-like mode, therefore, is $\Omega_- \approx \omega_\text{TO}$ (red curve), while the electronic plasmon is $\Omega_+ \approx \omega_{p\infty}$ (blue curve). At low carrier densities $(\omega_{p\infty} \ll \omega_\text{LO})$, the phonon-like mode is $\Omega_+ \approx \omega_\text{LO}$, while plasmon is $\Omega_- \approx \omega_{p0} \equiv \omega_{p\infty} \sqrt{\ve_\infty/\ve_0}$. In this regime the longitudinal oscillations are fast enough to screen plasma oscillations thereby modifying the plasmon mode from $\omega_{p\infty}$ to $\omega_{p0}$.}
\end{figure}

In the absence of magnetic field, the effects of electron-phonon coupling disappear ($\Pi_{ij}(\Omega_m)=0$ from Eqs.~\eqref{Pi}--\eqref{L}) and the only resonances present are phonon and plasmon ones. The eigenmodes of the system are then found from
\beq
\label{Z=0}
\begin{split}
\text{Det}\big[[\hat{\mathcal{D}}^0]^{-1}\big] =& - \Big( \frac{2\pi V}{\Omega_0^2} \Big)^3 \big( \Omega_m^2 + \omega_\text{TO}^2 \big)^2 \times \\
& \times \bigg[ \omega_\text{TO}^2 + \Omega_m^2 \bigg( 1 + \frac{\omega_{pi}^2}{\Omega_m^2 + \omega_{p\infty}^2} \bigg) \bigg]=0,
\end{split}
\eeq
where $\omega_\text{TO} \equiv \omega_\text{TO}(0, T)$ at $q=0$. Upon performing analytic continuation ($i\Omega_m \to \Omega + i0^+$) and solving Eq.~\er{Z=0} for $\Omega>0$, we get a double-degenerate solution at $\Omega = \omega_\text{TO}$, and two non-degenerate solutions $\Omega = \Omega_{\pm}$, where\cite{mahan:book,Ruhman:2016}
\beq
\label{pl-ph}
\Omega_{\pm} = \frac{1}{\sqrt{2}} \bigg[ \omega_{p\infty}^2 + \omega_\text{LO}^2 \pm \sqrt{-4\omega_{p\infty}^2 \omega_\text{TO}^2 + \big( \omega_{p\infty}^2 + \omega_\text{LO}^2 \big)^2} \bigg]^{1/2},
\eeq
with $\omega_\text{LO} \equiv \omega_\text{LO}(0) = \omega_\text{TO}^2 + \omega_{pi}^2$ as per Eq.~\er{LO}. The eigenvectors of $[\mathcal{D}^0]^{-1}$ suggest that the coupled phonon-plasmon modes \er{pl-ph}, as shown in Fig.~\ref{hyb_Z=0}, correspond to longitudinal polarization. Therefore, the coupling also occurs between plasmons and the longitudinal component of phonons. This makes sense because plasmons are longitudinal excitation in metals which is expected to couple only with LO modes.

The eigenmodes \er{pl-ph} can be understood in the asymptotic limits of high and low carrier densities $(n)$. At high $n$, $\omega_{p\infty} \gg \omega_\text{LO}$, and we get $\Omega_+ \approx \omega_{p\infty}$ and $\Omega_- \approx \omega_\text{TO}$. We can see that the phonon-like mode has a frequency $\omega_\text{TO}$, not $\omega_\text{LO}$; see Fig.~\ref{hyb_Z=0}. This is because the frequency difference between LO and TO modes is caused by long-ranged Coulomb interactions. At high $n$, or for $\omega_{p\infty} \gg \omega_\text{LO}$, the electron gas screens these long-ranged interactions and therefore the splitting $(\omega_{pi})$ between the LO and TO modes disappears. This is also consistent with our discussion in Sec.~\ref{distinction} that at high carrier densities the distinction between the energies of LO and TO modes disappears.

At low $n$, $\omega_{p\infty} \ll \omega_\text{LO}$ and the two roots are now $\Omega_+ \approx \omega_\text{LO}$ and $\Omega_- \approx \omega_{p\infty} \omega_\text{TO}/\omega_\text{LO}$. Using the Lyddane-Sachs-Teller (LST) relation,\cite{LST} one can write $\Omega_- \approx \omega_{p0}$, where $\omega_{p0}$ is the low energy plasma frequency 
\beq
\label{wp0}
\omega_{p0} = \sqrt{\frac{4\pi n e^2}{\ve_0 m_b}},
\eeq
\noindent which has static dielectric constant $\ve_0$ as the screening factor. So the plasmon-like mode is $\omega_{p0}$, not $\omega_{p\infty}$ as discussed in Sec.~\ref{distinction} and shown in Fig.~\ref{hyb_Z=0}. Physically this means that the longitudinal oscillations are fast enough to screen the plasma oscillations and modify $\omega_{p\infty}$ to $\omega_{p0}$. The coupling between optical modes and plasmons has been verified a long time ago in GaAs using Raman scattering and the result shows excellent agreement with the resonance frequencies obtained in Eq.~\er{pl-ph}.\cite{mooradian:1966}

In passing we note that as we go closer to the critical point, the low energy plasmon frequency $(\omega_{p0})$ is expected to vanish at the critical point, i.e., at $\omega_\text{TO} = 0$. Its contribution to the Coulomb interaction is, however, negligible as the value of the residue of the corresponding plasmon pole turns simultaneously to zero.

\subsubsection{Coupling between phonons, plasmons, and spin-flip modes, and their criticality}
\label{hybrid_B}
We now study the effect of magnetic field which splits electron band into spin-up and spin-down subbands. According to our model, it can affect phonons and underlying plasmons only through the electron-phonon interaction,  Eq.~\er{coupling}. The phonon self-energy (Eqs.~\eqref{Pi}--\eqref{L}) is now finite and therefore must be included to get the full phonon propagator \er{bareTO}. 

The eigenmodes of the full system can be obtained by solving $\text{Det}\big[ [\hat{\mathcal{D}}^0]^{-1} - \hat{\Pi} \big] = 0$ (after analytical continuation $i\Omega_m\to  \Omega+ i\delta$), which can be explicitly written as
\beq
\label{full}
\begin{split}
\text{Det} \bigg[ \big( \Omega_m^2 + \omega_\text{TO}^2 \big) \mathbb{1}_{\alpha\beta} &+ \frac{\omega_{pi}^2\Omega_m^2}{\Omega_m^2 + \omega_{p\infty}^2} \frac{q_\alpha q_\beta}{q^2} \\
&- \frac{\alpha L_Z}{15 \big( \Omega_m^2 + \Delta_Z^2 \big)} \tilde{\Pi}_{\alpha\beta} \bigg] = 0,
\end{split}
\eeq
where $\mathbb{1}_{\alpha\beta}$ is a component of $3\times3$ identity matrix,
\beq
\alpha \equiv \frac{\lambda^2 m_b^{5/2} \Omega_0^2}{2\pi^3}
\label{alpha}
\eeq
characterizes the strength of electron-phonon coupling \er{coupling} and, $L_Z$ and $\hat{\tilde{\Pi}}$ are given in Eqs.~\er{L} and \er{Pi_til}, respectively. 
The equation \er{full} leads to a polynomial equation of sixth order in $\Omega_m^2$ with six positive definite non-trivial roots for $\Omega$ ($i\Omega_m \to \Omega$). These correspond to hybridized electronic and phonon modes. The case of zero magnetic field leads to $\tilde{\Pi}=0$ (see Eq.~\eqref{Pi}) has been discussed in Sec. \ref{hybrid_Z=0}. There exists only four modes: two unhybridized TO modes and two hybridized modes arising due to the coupling of LO mode with plasmon.
The two additional modes that appear in finite magnetic field correspond to spin-flip transitions and will be discussed below in the regimes of high and low carrier densities, as outlined in Sec.~\ref{distinction}.

\begin{figure*}
\centering
\includegraphics[scale=0.53]{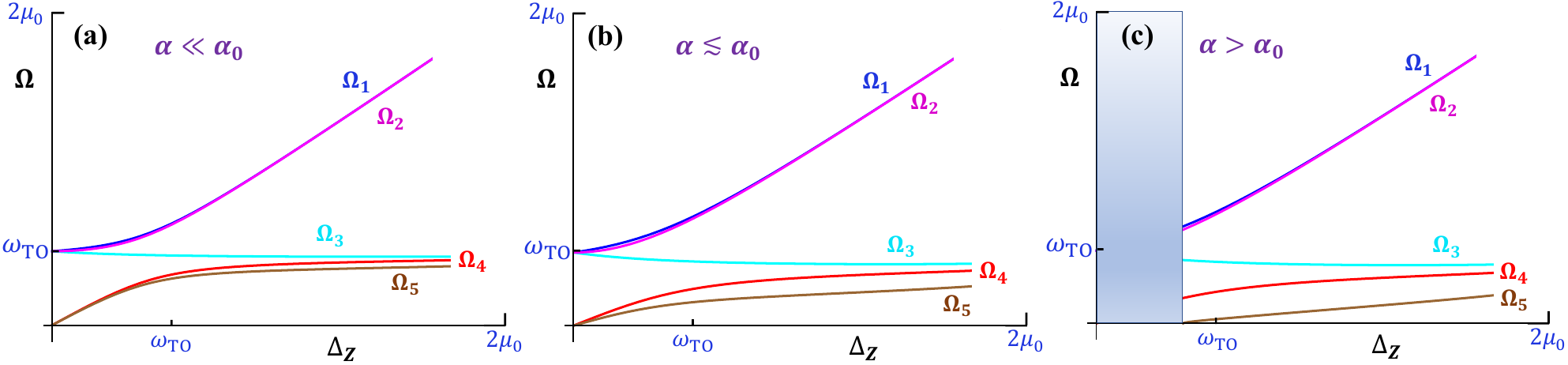}
\caption{\label{q=0} Evolution of $q=0$ eigenmodes in the high carrier density regime $(\omega_{p\infty} \gg \omega_{pi})$ as the coupling constant $\alpha$ \eqref{alpha} grows. The eigenmodes are plotted with $\Delta_Z$, where all the energy scales are assumed to be much smaller than $2\mu_0$. At low fields, there exist three phonon-like ($\Omega \approx \omega_\text{TO}$) and two spin-flip like ($\Omega \propto \Delta_Z$) modes. The electronic plasmon mode is at high energy and is not shown. (a) At small and finite $\alpha$, the coupling between phonons and spin-flip mode starts to take place, resulting in the hybridization of modes. (b) The lowest energy mode $\Omega_5$ (brown) is stable as long as $\alpha < \alpha_0$, and starts to become unstable as the critical point $\alpha = \alpha_0$ is approached. (c) For $\alpha>\alpha_0$, the lowest energy mode $\Omega_5$ is unstable at low fields, indicating a phase transition into the polar phase (shaded region).} 
\end{figure*}

\paragraph{High carrier density metals: $\Omega, \omega_{pi} \ll \omega_{p\infty}$} \label{ws<wp} 
Here we consider the case of a high-density metals characterized by a large bare plasma frequency, $\Omega, \omega_{pi} \ll \omega_{p\infty}$. Then, as discussed in Sec.~\ref{distinction}, the LO-TO splitting can be ignored (middle term in brackets in Eq.~\er{full}). In Fig.~\ref{q=0} we plot the resulting solutions of Eq.~\er{full} as a function of the Zeeman splitting $\Delta_Z$ and the coupling $\alpha$ \eqref{alpha} (panels (a) -(c)). 

One observes that there exists five solutions, corresponding to three phonon modes hybridizing with two spin-flip modes. Since for high carrier density metals $\omega_{p\infty}$ is large, the corresponding plasmon-like mode exists at large energies and therefore is excluded from this discussion. It, however, shows up in the low carrier density metals which we will discuss in Sec.~\ref{ws>wp}. 

\textit{Weak field regime:} We first focus on low fields $\Delta_Z \ll \omega_\text{TO}$ and sufficiently low $\alpha$ (Fig.~\ref{q=0}(a)), where all eigenvalues are real. The eigenfrequencies are given by
\beq
\label{small_Z}
\begin{split}
\Omega_1 &\approx \omega_\text{TO} + \Delta_Z \frac{\alpha (2\mu_0)^{3/2}}{6\omega_\text{TO}^2} + \Delta_Z^2 \frac{\alpha(2\mu_0)^{3/2}}{6\omega_\text{TO}^3} \bigg( 1 - \frac{\alpha(2\mu_0)^{3/2}}{4\omega_\text{TO}^2} \bigg) \\
&\hspace{5cm} + \mathcal{O}(\Delta_Z^3), \\
\Omega_2 &\approx \omega_\text{TO} + \Delta_Z^2 \frac{\alpha(2\mu_0)^{3/2}}{6\omega_\text{TO}^3} + \mathcal{O}(\Delta_Z^4), \\
\Omega_3 &\approx \omega_\text{TO} - \Delta_Z \frac{\alpha (2\mu_0)^{3/2}}{6\omega_\text{TO}^2} + \Delta_Z^2 \frac{\alpha(2\mu_0)^{3/2}}{6\omega_\text{TO}^3} \bigg( 1 - \frac{\alpha(2\mu_0)^{3/2}}{4\omega_\text{TO}^2} \bigg) \\
&\hspace{5cm} + \mathcal{O}(\Delta_Z^3), \\
\Omega_4 &\approx \Delta_Z \Big| 1 - \frac{\alpha (2\mu_0)^{3/2}}{3\omega_\text{TO}^2} \Big| + \mathcal{O}(\Delta_Z^3), \\
\Omega_5 &\approx \Delta_Z \sqrt{1 - \frac{2 \alpha (2\mu_0)^{3/2}}{3\omega_\text{TO}^2}} + \mathcal{O}(\Delta_Z^3),
\end{split}
\eeq
where $\mu_0$ is the chemical potential in the absence of magnetic field. We identify two modes ($\Omega_{4,5}$ in Eq.~\er{small_Z}) with frequency vanishing linearly with $\Delta_Z$, and three modes ($\Omega_{1,2,3}$ in Eq.~\er{small_Z}) saturating at a finite energy $\omega_\text{TO}$ as $\Delta_Z$ approaches zero, as shown in Fig.~\ref{q=0}(a). The latter three modes, $\Omega_{1,2,3}$, can then be identified as phonon modes which are degenerate in the absence of magnetic fields. $\Omega_{4,5}$, on the other hand, correspond to the electronic spin-flip transitions, with renormalized $g$-factor due to electron-phonon coupling, between the spin-split subbands of the conduction band. One can clearly see that at $\alpha\to0$, the effective $g$-factor for $\Omega_{4,5}$ approaches its bare value. For the particular case of $\alpha=0$, however, Eq. \er{full} does give rise to only four roots, with $\Omega_{4,5}$ being absent. This describes the decoupling of the spin-flip resonances from the phonon response in the absence of coupling \eqref{coupling}. 

Already at small fields, the degeneracy of modes stemming from phonons ($\Omega_1$, $\Omega_2$ and $\Omega_3$) and electrons ($\Omega_4$ and $\Omega_5$) breaks, which deserves explanation. If we assume that there is no electron-phonon interaction, i.e., the corresponding coupling constant, parameterized by $\alpha$, is zero, then the precession-like spin-flip resonances that occur in any doped semiconductor with fields applied along, say, $z$-direction, are double degenerate ($\Omega_4 = \Omega_5 = \Delta_Z$ at $\alpha=0$ in Eq.~\er{small_Z}) as a result of rotational symmetry in the $xy$-plane. Likewise, there is no distinction between two transverse components of the phonon modes due to underlying rotational symmetry in the $xy$-plane. The absence of LO-TO splitting for high carrier density metals further results in triple degenerate phonon modes ($\Omega_1 = \Omega_2 = \Omega_3$ at $\alpha=0$ in Eq.~\er{small_Z}). 

In the presence of interaction, i.e., $\alpha \neq 0$, and at weak magnetic fields, the modes are expected to couple and break degeneracy too. The interaction Hamiltonian \er{coupling} is invariant under simultaneous rotation of electron spin, momenta and phonon polarization. The magnetic field, however, breaks the spin-rotation symmetry for the conduction electrons and, consequently for the phonon polarization due to the interaction \er{coupling}. This explains qualitatively the lifting of degeneracies at finite fields in Eq.~\er{small_Z} and Fig.~\ref{q=0}(a).

\textit{Softening of spin-flip mode}: 
On increasing $\alpha$, the slope of the spin-flip modes at low $\Delta_Z$ decreases, as shown in Fig.~\ref{q=0}(b). Moreover, for $\alpha$ larger than a critical value, the frequency of the lowest-lying eigenmode ($\Omega_5$) becomes imaginary at low magnetic field, as shown in Fig.~\ref{q=0}(c), indicating the instability of the system.
Beyond that point, a phase transition into the polar phase occurs and the calculations performed here need to be modified (see also Appendix \ref{app:pert}).
As the instability takes place first at  $\Delta_Z=0$, it is instructive to consider eigenfrequencies at small fields given in Eq.~\er{small_Z} at $\alpha\leq \alpha_0$.
At small fields the spin-flip modes ($\Omega_4$ and $\Omega_5$) disperse linearly with $\Delta_Z$, with effective $g$-factor as a function of coupling constant ($\alpha$), TO mode frequency ($\omega_\text{TO}$), and carrier density $n$ (through $\mu_0$).
One observes that the frequency of one of the spin-flip modes, $\Omega_5$, becomes imaginary for $\alpha>\alpha_0$, where 
\beq
\label{a0}
\alpha_0 = \frac{3\omega_\text{TO}^2}{2(2\mu_0)^{3/2}}.
\eeq 
From experimental point of view, the parameter $\alpha$ is fixed and the transition is really driven by the softening of $\omega_\text{TO}$. Eq.~\eqref{a0} can then be viewed as a condition for $\omega_\text{TO}$. Using Eq.~\er{alpha}, the instability condition, $\alpha>\alpha_0$, translates to
\beq
\label{crit:TO}
\omega_\text{TO} < \lambda\Omega_0 \sqrt{\frac{n m_b}{\pi}}.
\eeq
Additionally, $\Omega_5$ can be rewritten as
\beq
\label{mode5a}
\Omega_5 \approx g\mu_B \bigg( 1 - \frac{\alpha}{\alpha_0} \bigg)^{1/2} B,
\eeq
where $\alpha_0$ is defined in Eq.~\er{a0}, which shows a linear dispersion with magnetic field $B$, with the effective $g$-factor, $g^* = g(1-\alpha/\alpha_0)^{1/2}$, which softens at $\alpha=\alpha_0$. For $\alpha>\alpha_0$, $g^*$ becomes imaginary, making $\Omega_5$ also imaginary, indicating the onset of instability of the system. The form of $\Omega_4$ in Eq.~\er{small_Z} suggests that the $g$-factor for $\Omega_4$ also vanishes at $\alpha = 2\alpha_0$. However, unlike $\Omega_5$ it never becomes imaginary, suggesting that only the softening of $\Omega_5$ indicates the instability.

The dependence of $\alpha_0$ on $\omega_\text{TO}$ \er{a0} and the instability condition \er{crit:TO} demonstrates the direct impact of spin-orbit mediated interaction \er{coupling} between conduction electrons and TO modes, in agreement with previous works.\cite{Maria:2020, kanasugi:2018, kanasugi:2019, Maria:review} 
This is in contrast to the gradient-like electron-phonon coupling which vanishes in the case of transverse phonon mode. As a result, the soft TO mode frequency effectively decouples from the electronic states for an isotropic system.\cite{Maria:2020, volkov2021, kumar:2021}

Thus we have demonstrated that the signature of the instability of the system in the $q=0$ regime is the softening of the spin-flip electronic mode $\Omega_5$ at weak fields. Note that this does not imply the instability to be of the electronic character as we have implicitly considered finite frequencies here due to the condition $\Omega \gg v_F q$. In Sec.~\ref{finite_q}, we will show that the phononic character of the instability is recovered in the $\Omega=0$ and $q\to0$ limit, which indicates the actual polar instability of the system.

\textit{Moderate and large fields:} We now propose a way to obtain the coupling constant $\lambda$. At $\alpha \ll \alpha_0$, when the system is stable, the degeneracy of phonons and spin-flip modes breaks, as shown in Fig.~\ref{q=0}(a). As the magnetic field is increased, the interaction between phonons and spin-flip resonances occurs, resulting in the avoided crossing (anticrossing) of modes at $\Delta_Z \sim \omega_\text{TO}$. The energy splitting near the anticrossing point can be obtained analytically. From Fig.~\ref{q=0}(a), one observes that the maximum splitting near the anticrossing point occurs between $\Omega_5$ and, $\Omega_1$ or $\Omega_2$. As the splitting between $\Omega_1$ or $\Omega_2$, as evident from Fig.~\ref{q=0}, is very small, we focus on the splitting between $\Omega_2$ and $\Omega_5$. The exact analytical forms of $\Omega_2$ and $\Omega_5$ are given by
\beq
\label{w25}
\Omega_{2(5)} = \frac{1}{\sqrt{2}} \bigg[ \omega_\text{TO}^2 + \Delta_Z^2 \pm \sqrt{(\omega_\text{TO}^2 - \Delta_Z^2)^2 + \frac{8\alpha}{15} \Delta_Z L_Z} \bigg]^{1/2},
\eeq
with $\Omega_2$ corresponding to the one with `+' sign, while $\Omega_5$ to the one with `-' sign. Here $\alpha$ and $L_Z \equiv L\big[ \mu(\Delta_Z), \Delta_Z \big]$ are defined in Eqs.~\er{alpha} and \er{L}, respectively. 

The splitting between square of energies is given by
\beq
\label{diff}
\Delta\Omega^2 \equiv \Omega_2^2 - \Omega_5^2 = \sqrt{(\omega_\text{TO}^2 - \Delta_Z^2)^2 + \frac{8\alpha}{15} \Delta_Z L_Z}.
\eeq
We observe that the splitting depends strongly on the carrier density through $L_Z$: up to leading order in $\Delta_Z$, $L_Z \approx 5\Delta_Z (2\mu_0)^{3/2} \propto n$, where $n$ is free carrier density. This implies that the splitting between square of energies increases with the carrier density as $\Delta\Omega^2 \propto n^{1/2}$, assuming $\alpha$ remains independent from $n$. The form of $\alpha$ can be obtained from Eq.~\er{diff} as
\beq
\label{a1}
\alpha = \frac{3}{8\Delta_Z^2 (2\mu_0)^{3/2}} \Big[ \big( \Delta\Omega^2 \big)^2 - \big( \omega_\text{TO}^2 - \Delta_Z^2 \big)^2 \Big].
\eeq

Near the anticrossing point ($\Delta_Z \approx \omega_\text{TO}$), the second term of Eq.~\er{a1} can be ignored. Then using Eq.~\er{alpha}, we find the coupling constant $\lambda$ to be
\beq
\label{lambda}
\lambda = \sqrt{\frac{\pi (\Delta\Omega^2)^2}{4 n m_b\Delta_Z^2 \Omega_0^2}},
\eeq
where $n$ is carrier density, $m_b$ is effective mass of electrons and $\Omega_0 \equiv \sqrt{\ve_\infty} \omega_{pi}$. $\Omega_0$ can be obtained experimentally for doped polar metals from measurements on the parent insulator, but may not be straightforward to extract for intrinsic polar metals. On the other hand, the definition of phonon Green's function, Eqs.~\er{full} and \er{alpha} suggests that the expansion parameter for the electron-phonon interaction is really $\lambda \Omega_0$. As follows from \eqref{lambda}, from the measured $\Delta\Omega^2$ at a given carrier density and the strength of the applied magnetic field, one can extract $\lambda\Omega_0$.

Approaching the polar critical point from the paraelectric phase, $\omega_\text{TO} \equiv \omega_\text{TO}(0, T)$ softens to a value that is small in the weak-coupling case (see Eq. \eqref{a0} and the related discussion). It follows then that the required magnetic field to extract $\lambda$, $B \approx \omega_\text{TO}/g\mu_B$, also would be small enough and practically attainable; see also a detailed discussion in Sec.~\ref{exp}.

Finally, at magnetic fields large compared to $\omega_\text{TO}$ but still smaller than $2\mu_0$, such that $\omega_\text{TO}\ll\Delta_Z\ll2\mu_0$, the spin-flip modes ($\Omega_4$ and $\Omega_5$) evolve into phonon modes, while $\Omega_3$ remains phononic. It can be shown analytically that in this limit their frequencies are shifted downwards by a field-independent amount from $\omega_\text{TO}$ due to coupling to electrons. Lastly, $\Omega_1$ and $\Omega_2$ which started off as phonon modes evolve into spin-flip modes at large fields and their frequencies vary linearly with $\Delta_Z$. The evolution of eigenmodes of mixed character at all fields but small $\alpha$ is shown in Fig.~\ref{q=0}(a).

\begin{figure}
\centering
\includegraphics[scale=0.35]{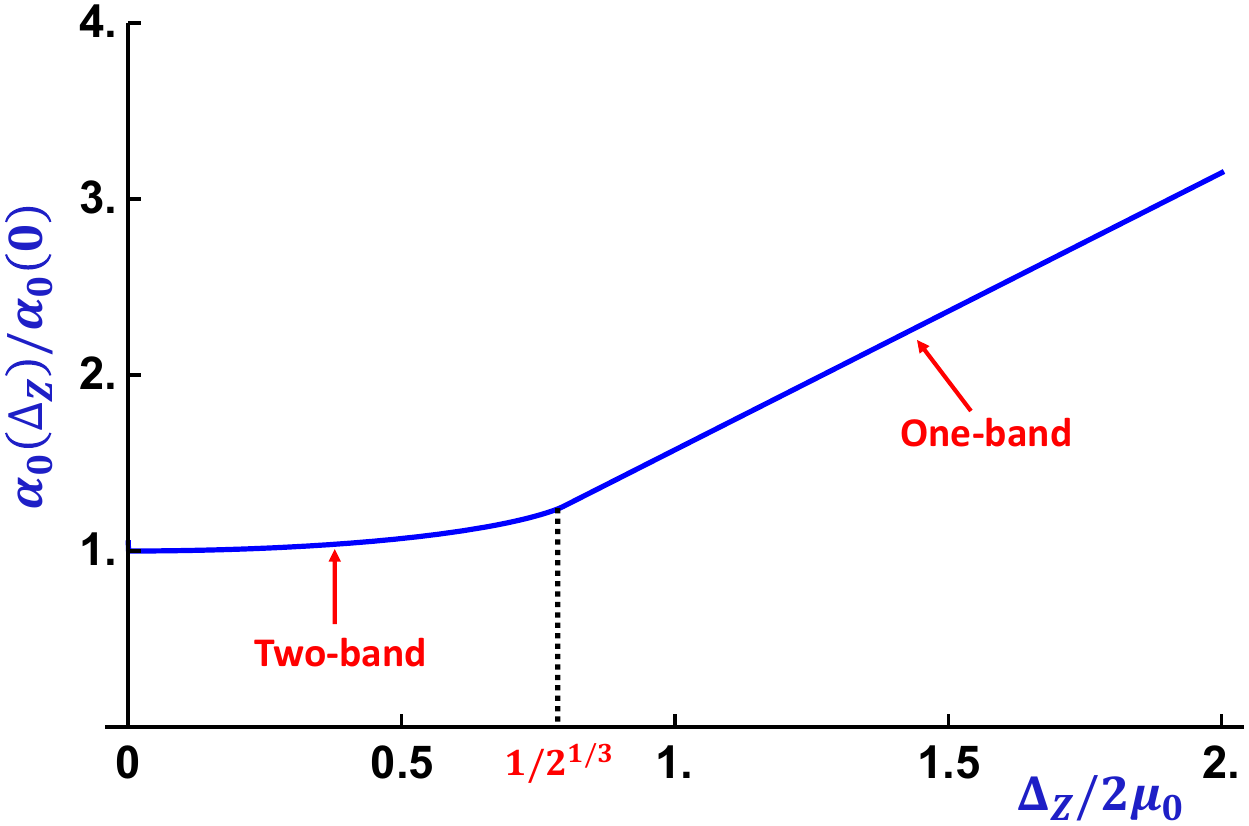}
\caption{\label{phase} Phase diagram of $\alpha(\Delta_Z)/\alpha(0)$ vs $\Delta_Z/2\mu_0$, where $\alpha(0) \equiv \alpha_0$. It can be seen clearly that with the increase in $\Delta_Z$, the system goes from two-band case (both spin-up and spin-down subbands are partially occupied) to one-band case (only spin-down subband is partially occupied). The crossover happens at $\Delta_Z = 2\mu_0/2^{1/3}$.}
\end{figure}

\textit{Polar instability at $\Delta_Z\neq0$ and phase diagram}:  
The critical coupling constant $\alpha_0$ \er{a0} was obtained at small fields based on our observation in Fig.~\ref{q=0}(c) that the instability starts to take place at $\Delta_Z=0$. Now we will determine how the instability condition Fig.~\ref{q=0}(c) is modified by the magnetic field and obtain a phase diagram for the critical coupling constant $\alpha_0(\Delta_Z)$ as a function of magnetic field. We do not limit the range of magnetic field and allow the corresponding energy scale $\Delta_Z$ to exceed the Fermi energy ($2\mu_0$ to be precise) to reach the one-band limit when the system is fully polarized. The corresponding phase diagram, obtained numerically, is plotted in Fig.~\ref{phase}. 

We observe in Fig.~\ref{phase} that the critical coupling constant starts off with the value $\alpha_0$ at $\Delta_Z=0$ and increases with the magnetic field. In the two-band regime, it increases slowly with field, whereas in the one-band regime the critical coupling constant increases faster following a linear dependence. The increase of  $\alpha_0(\Delta_Z)$ conversely implies that the critical value of $\omega_\text{TO}$ becomes smaller with increasing field. The crossover from two-band to one-band regime happens at $\Delta_Z = 2^{-1/3}(2\mu_0)$; 
see Eq.~\er{chem_tot} and Fig.~\ref{chem_pot} of Appendix~\ref{app:chem_3D} for the variation of chemical potential with magnetic fields.

To understand the main features of Fig.~\ref{phase} analytically, we use the stability condition of mode $\Omega_5$ for the two-band case \er{w25}, written as
\beq
\label{cond1}
\sqrt{(\omega_\text{TO}^2 - \Delta_Z^2)^2 + \frac{8\alpha}{15} \Delta_Z L(\mu, \Delta_Z)} < \omega_\text{TO}^2 + \Delta_Z^2,
\eeq
and consider higher order magnetic field corrections to $L_Z$ \er{L}. We remind the readers that $L_Z$ is actually a functional of chemical potential which is a function of magnetic field \er{chem_2band}; while taking higher order field corrections for $L_Z$, this must be kept in mind. The Eq.~\er{cond1} leads to the condition on the coupling constant $\alpha$: 
\beq
\label{a<a0}
\alpha < \frac{15 \omega_\text{TO}^2 \Delta_Z}{2L_Z}.
\eeq
This condition is of course for the two-band case as is evident from the form of $L_Z$ \er{L}. Collecting higher order magnetic field corrections of $L_Z$ in Eq.~\er{a<a0}, the stability constraint for $\Omega_5$ can be extended for finite fields as $\alpha < \alpha_0(\Delta_Z)$, where
\beq
\label{coup_2band}
\frac{\alpha_0(\Delta_Z)}{\alpha_0(0)} = \bigg[ 1 + \frac{1}{4} \bigg( \frac{\Delta_Z}{2\mu_0} \bigg)^2 + \frac{9}{80} \bigg( \frac{\Delta_Z}{2\mu_0} \bigg)^4  + \mathcal{O} \bigg( \frac{\Delta_Z}{2\mu_0} \bigg)^6 \bigg],
\eeq
with $\alpha_0(0) \equiv \alpha_0$ obtained in Eq.~\er{a0}.

For fixed carrier density, as the magnetic field is increased further, the system becomes fully polarized, i.e., only spin-down subband remains (partially) occupied. In this case, the one-band analogue of Eq.~\er{w25} for $\Omega_5$ takes the form
\beq
\label{mode5_1band}
\Omega_5 = \frac{1}{\sqrt{2}} \Bigg[ \omega_\text{TO}^2 + \Delta_Z^2 - \sqrt{(\omega_\text{TO}^2 - \Delta_Z^2)^2 + \frac{8\alpha}{15} \Delta_Z L_Z^\text{OB}} \Bigg]^{1/2},
\eeq
where $L_Z^\text{OB} = \big( 2\mu(\Delta_Z) + \Delta_Z \big)^{5/2}$ is the one-band analogue of $L_Z$, with $\mu(\Delta_Z)$ as the field dependent chemical potential obtained in Eq.~\er{chem_1band} for the one-band case. Unlike the two-band case, the exact solution for the chemical potential as a function of field is possible in the one-band limit \er{chem_1band}, substituting which in Eq.~\er{mode5_1band} gives the stability condition for $\Omega_5$ as $\alpha < \alpha_0(\Delta_Z)$, where
\beq
\label{coup_1band}
\frac{\alpha_0(\Delta_Z)}{\alpha_0(0)} = \frac{5}{2^{5/3}} \Big( \frac{\Delta_Z}{2\mu_0} \Big).
\eeq

The above allows to understand the phase diagram shown in Fig.~\ref{phase}, in the limit of small and large $\Delta_Z$ (with respect to $\mu_0$).

\begin{figure*}
\centering
\includegraphics[scale=0.53]{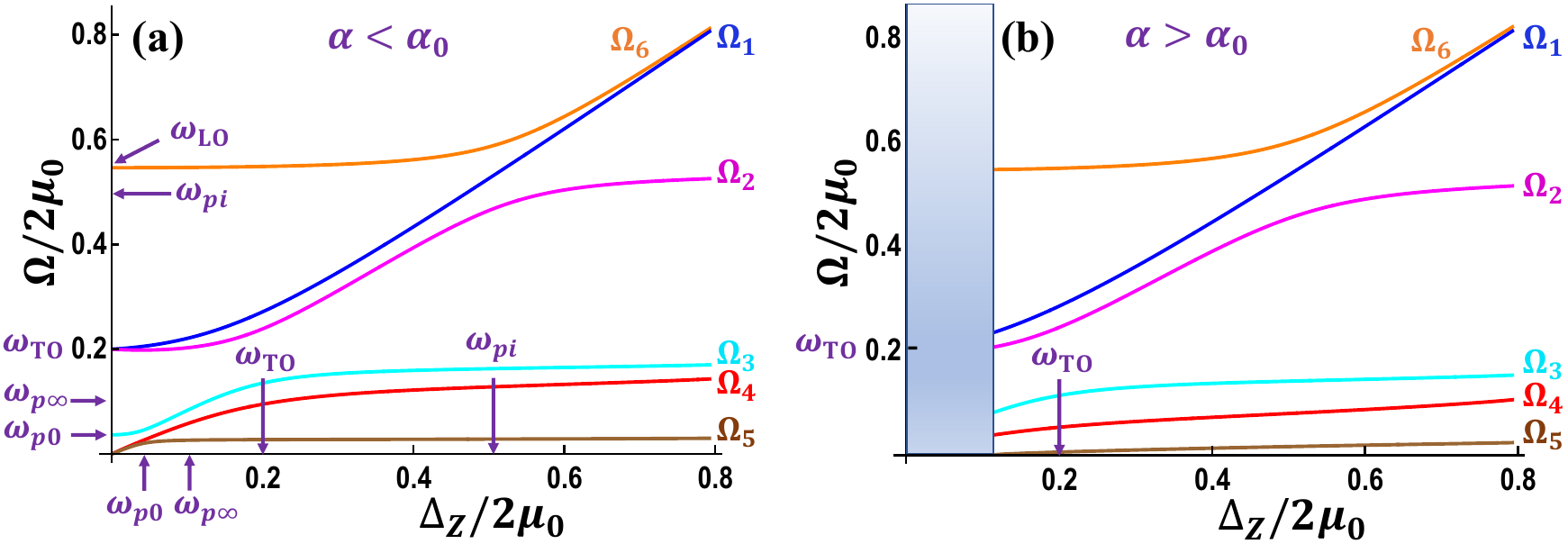}
\caption{\label{split_q=0} Hybridization of modes at $q=0$ in the low carrier density regime $(\omega_{p\infty} \ll \omega_{pi})$ where the LO-TO splitting effects are significant. In the plot, all energy scales are in units of $2\mu_0$ such that $\omega_{pi}/2\mu_0=0.5$, $\omega_\text{TO}/2\mu_0=0.2$ and $\omega_{p\infty}/2\mu_0=0.1$. The chosen value of $\omega_\text{TO}$ corresponds to $\alpha_0=0.06$, in units of $\sqrt{2\mu_0}$. The polarization angle $\theta_\bq$ has been arbitrarily chosen to be $\pi/3$. (a) Hybridization of modes in the presence of magnetic fields for $\alpha<\alpha_0$. The lowest energy mode $\Omega_5$ is stable. (b) For $\alpha>\alpha_0$, $\Omega_5$ becomes imaginary which indicates the onset of phase transition.}
\end{figure*}

\paragraph{Low-density metals: $\omega_{p\infty} \ll \omega_{pi}$}\label{ws>wp} 
We now turn to the case of low densities where  $\omega_{p\infty} \ll \omega_{pi}$, appropriate for doped ferroelectric or paraelectrics, such as BaTiO$_3$\cite{kolodiazhnyi:2010, takahashi:2017}, or SrTiO$_3$\cite{Rischau2017, Engelmayer2019b, Wang2019}. The LO-TO splitting effects become large in this regime and we must take the middle term of Eq.~\er{full} into account. In addition to that, the electronic plasma frequency is of the order of phonon frequency in this regime. Consequently, the electronic plasmon and its interaction with the spin-flip and phonon modes will play an important role.
The eigenmodes are shown in Fig.~\ref{split_q=0}(a).
As expected, there are six positive definite roots in this regime corresponding to phonon-plasmon modes coupled with spin-flip modes. The coupling between modes is characterized by anticrossing in several magnetic field regimes. In Fig.~\ref{split_q=0} the parameters are chosen such that $\omega_{p\infty} \ll \omega_\text{TO} \ll \omega_{pi}$. This also implies $\omega_\text{TO} \ll \omega_\text{LO}$, according to Eq.~\er{LO}. Like in the high carrier density regime, discussed in Sec.~\ref{ws<wp}, in this regime also one of the spin-flip modes $(\Omega_5)$ becomes unstable (imaginary) above a certain value of $\alpha$, as shown in Fig.~\ref{split_q=0}(b). Moreover, the instability occurs at $\Delta_Z = 0$, same way as in Sec.~\ref{ws<wp}. We study the weak field regime to obtain the critical value of $\alpha$ for this instability followed by moderate and large field regimes to study mode anticrossing.

\textit{Weak magnetic fields}: At small fields, the eigenfrequencies are obtained as
\beq
\label{full_small_Z}
\begin{split}
\Omega_1 &\approx \omega_\text{TO} + \Delta_Z \frac{\alpha (2\mu_0)^{3/2} |\cos\theta_\bq|}{6\omega_\text{TO}^2} + \mathcal{O}(\Delta_Z^2), \\
\Omega_2 &\approx \omega_\text{TO} - \Delta_Z \frac{\alpha (2\mu_0)^{3/2} |\cos\theta_\bq|}{6\omega_\text{TO}^2} + \mathcal{O}(\Delta_Z^2), \\
\Omega_3 &\approx  \frac{1}{\sqrt{2}} \bigg[ \omega_{p\infty}^2 + \omega_\text{LO}^2 - \sqrt{-4\omega_{p\infty}^2 \omega_\text{TO}^2 + \big( \omega_{p\infty}^2 + \omega_\text{LO}^2 \big)^2} \bigg]^{1/2} \\
& \hspace{5cm} + \mathcal{O}(\Delta_Z^2), \\
\Omega_4 &\approx \Delta_Z \Big| 1 - \frac{\alpha (2\mu_0)^{3/2}}{3\omega_\text{TO}^2} \Big| + \mathcal{O}(\Delta_Z^2), \\
\Omega_5 &\approx \Delta_Z \sqrt{1 - \frac{2 \alpha (2\mu_0)^{3/2}}{3\omega_\text{TO}^2}} + \mathcal{O}(\Delta_Z^2) \\
\Omega_6 &\approx  \frac{1}{\sqrt{2}} \bigg[ \omega_{p\infty}^2 + \omega_\text{LO}^2 + \sqrt{-4\omega_{p\infty}^2 \omega_\text{TO}^2 + \big( \omega_{p\infty}^2 + \omega_\text{LO}^2 \big)^2} \bigg]^{1/2} \\
& \hspace{5cm} + \mathcal{O}(\Delta_Z^2),
\end{split}
\eeq
where $\theta_\bq$ is the polar angle of $\bq$ with respect to the $z$-axis. The effect of the direction of $\bq$ results from the combination of LO-TO splitting and Zeeman field. For $\bq$ along $z$ axis ($\theta_\bq=0$), $\Omega_{1,2}$ can be observed to coincide with $\Omega_{1,3}$ in the high-density limit (see Eq. \eqref{small_Z}). These modes correspond to phonons polarized in the $xy$ plane, and thus LO-TO splitting does not affect them for $\bq$ along $z$. The splitting $\sim \Delta_Z$ between them is caused by the off-diagonal terms in the phonon self-energy \eqref{Pi_til}, that affects only phonons, polarized in $xy$ plane. As $\theta_\bq$ increases, one of the transverse modes' polarization has to rotate towards $z$, such that for $\theta_\bq= \pi/2$ it is fully along $z$. In that case, the self-energy \eqref{Pi_til} does not lead to direct hybridization of the two transverse modes (since there are no off-diagonal elements coupling $z$-polarized eigenstates to the other ones), explaining the absence of a $\sim \Delta_Z$ splitting between $\Omega_{1,2}$ at $\theta_\bq= \pi/2$. Thus teh results of the previous section can only be recovered in the $\omega_{pi} \ll \Delta_Z$ limit, where the LO-TO splitting can be ignored.

At small fields, the coupling between different modes stemming from phonons, plasmons and spin-flips is weak. 
The eigenmodes shown in Fig.~\ref{split_q=0}(a) can be understood in a following manner. In the $\omega_{p\infty} \ll \omega_\text{TO} \ll \omega_{pi}$ regime, $\Omega_4$ and $\Omega_5$ are spin-flip modes and disperse linearly with $\Delta_Z$. Mode $\Omega_3$ corresponds to plasmon with $\ve_0$ as a screening factor. And finally $\Omega_1$ and $\Omega_2$ correspond to TO modes, while $\Omega_3$ to LO mode. 
This allows to understand the instability of one of the spin-flip modes $\Omega_5$ in the same way as it is discussed in Sec.~\ref{ws<wp}. Indeed, the form of $\Omega_5$ in Eq.~\er{full_small_Z} tells us that the mode becomes unstable for $\alpha > \alpha_0$, where $\alpha_0$ is defined in Eq.~\er{a0} which depends on TO mode frequency.
This is the same condition as that obtained for $\Omega_5$ in the high carrier density ($\omega_{p\infty} \gg \omega_{pi}$) regime.

At  $\Delta_Z=0$, one can find a clear distinction between the double-degenerate TO mode ($\Omega_1$ and $\Omega_2$ in Eq.~\er{full_small_Z} and Fig.~\ref{split_q=0}(a)) and the longitudinal modes ($\Omega_3$ and $\Omega_6$ in Eq.~\er{full_small_Z} and Fig.~\ref{split_q=0}(a)), arising from the hybridization between the electronic plasmon and the LO phonon. Moreover, as long as $\omega_{p\infty} \ll \omega_{pi}$ condition holds, the hybridization between plasmon and LO phonon is weak and therefore $\Omega_3$ and $\Omega_6$ can be identified as plasmon and LO phonon, respectively. As discussed in Sec.~\ref{hybrid_Z=0}, the plasmon frequency is $\omega_{p0}$ in this case, due to screening by fast longitudinal oscillations.

Overall, we have demonstrated that in the presence of LO-TO splitting and low-energy electronic plasmon, the polar instability of the system in the $q=0$ regime manifests itself in the softening of the spin-current electronic mode $\Omega_5$.

\textit{Moderate and large magnetic fields}: As the magnetic field is increased, a coupling of spin-flip modes with plasma and, TO and LO modes occurs, which is characterized by anticrossings at $\Delta_Z \sim \omega_{p0}$, $\Delta_Z \sim \omega_\text{TO}$ and $\Delta_Z \sim \omega_\text{LO}$, respectively. We reiterate that in the low carrier density regime, the frequency of the plasmon mode is $\omega_{p0}$, not $\omega_{p\infty}$. Therefore, the strong hybridization of spin-flip modes with plasmons occurs first at $\Delta_Z \sim \omega_{p0}$, and then with TO modes at $\Delta_Z \sim \omega_\text{TO}$.
As the magnetic field increases further, the LO mode and one of the spin-flip modes hybridize at $\Delta_Z \sim \omega_\text{LO}$. Due to this non-trivial hybridization at moderate and large fields, $\Omega_4$ and $\Omega_5$ which started off as spin-flip modes, evolve into one of the TO modes and low energy plasma mode, respectively. 
Mode $\Omega_3$ which started off as $\omega_{p0}$ evolves into another TO mode, while $\Omega_2$ transforms from one of the TO modes into the LO mode. Finally, $\Omega_1$ and $\Omega_6$ transform from corresponding TO and LO modes to spin-flip modes at large fields.

\section{Properties of the polar transition}
\label{finite_q}
In the previous section, we discussed the excitations of the spin-orbit coupled polar metal for $\Omega \gg v_Fq$ regime. One of our findings was the instability of the low-energy spin-flip mode above a critical value of the coupling constant (or for TO phonon frequency below a critical value \er{crit:TO}). In particular, its frequency depends linearly on the applied field $\Delta_Z$, but the slope of this dependence (effective $g$ -factor) vanishes at $\alpha = \alpha_0$. However, to assess the thermodynamic behavior of the system on the approach to the (quantum or classical) transition, we need to explore $\Omega \ll q^2/2m_b \ll v_Fq$ regime, where the static limit $\Omega=0$ can be studied.

The goals of this section are elucidate the nature of the soft mode: to understand whether it occurs at $q \to 0$ finite $q$ and what is its polarization (assuming it has phononic character). 
To study this we calculate phonon self-energy, $\Pi_{\alpha\beta}$, up to terms of order $q^2$ assuming $q^2/2m_b \ll v_F q \ll \Delta_Z$, and evaluate the phonon eigenenergies and eigenvectors in this limit.

\subsection{Self-energy at $\Omega=0$ and finite $q$}
We calculate self-energy at $\Omega=0$ in the limit $q^2/2m_b \ll v_F q \ll \Delta_Z$. We put $\Omega=0$ in the general expression for self-energy \er{se1} to get
\beq
\label{se_q}
\begin{split}
\Pi_{\alpha\beta} (\bq) = \frac{\lambda^2}{2} & \sum_{s \bar{s}} \int_\bk k^2 f_{\alpha\beta}^{s \bar{s}}(\theta, \phi) \times \\
& \times \frac{n_F(\xi_{\bk-\bq/2}^{\bar{s}}) - n_F(\xi_{\bk+\bq/2}^s)}{- \big( \ve_{\bk+\frac{\bq}{2}} - \ve_{\bk-\frac{\bq}{2}} \big) - (s-\bar{s})\frac{\Delta_Z}{2}},
\end{split}
\eeq
where the factors $f_{\alpha\beta}^{s \bar{s}}$ are given in Eq.~\er{coh_fac}. Note that in contrast to $\Omega\gg v_Fq$ case, the intraband contributions to $\Pi_{xx}$, $\Pi_{yy}$ and $\Pi_{xy}$ are not required to vanish by the conservation law for the spin current component $j_i^{\sigma_z}$. On the other hand, $\Pi_{zz}$ is not affected by this, as it consists only of $j_i^{\sigma_{x,y}}$ correlators, leading to interband transitions.

Replacing the Fermi functions in Eq.~\er{se_q} by $\Theta$-functions and upon writing $\bk \cdot \bq = k \bar{q}$, with $\bar{q} = q_x \sin\theta_\bk \cos\phi_\bk + q_y \sin\theta_\bk \sin\phi_\bk + q_z \cos\theta_\bk$, we obtain the conditions on the limits of $k$-integral. Upon using these conditions for $k$-integration, we obtain self-energy at $q\to0$ and up to $\Delta_Z^2$ order:
\beq
\label{Pi_q<Z}
\begin{split}
\Pi_{\alpha\beta}(\bq) \approx & - \frac{2\pi V}{\Omega_0^2}\frac{\alpha (2\mu_0)^{3/2}}{4} \Bigg[ \bigg( \frac{8}{3} - \frac{q^2}{3k_F^2} \bigg) \delta_{\alpha\beta} - \frac{q_\alpha q_\beta}{3k_F^2} \\
& - \frac{1}{3} \bigg( \frac{\Delta_Z}{2\mu_0} \bigg)^2 (\delta_{\alpha\beta} + \delta_{\alpha 3}) + \bigg( \frac{\Delta_Z}{2\mu_0} \bigg)^2 \bigg\{ \frac{q_\alpha q_\beta}{12k_F^2} \\
&\,\,\,\,\, + \delta_{\alpha\beta} \bigg( \frac{17q^2 - 6q_z^2}{480k_F^2} - \delta_{\alpha3} \frac{29q^2 - 12q_z^2}{480k_F^2} \bigg) \\
& + (\delta_{\alpha 1} \delta_{\beta 3} + \delta_{\alpha 3} \delta_{\beta 1} + \delta_{\alpha 2} \delta_{\beta 3} + \delta_{\alpha 3} \delta_{\beta 2}) \frac{q_\alpha q_\beta}{80k_F^2} \bigg\} \Bigg].
\end{split}
\eeq
Terms of order $\mathcal{O}((q^2/2m_b)^2/2\mu_0\Delta_Z)$ and higher are ignored. One observes that in contrast to the $\Omega \gg v_F q$ case (see Eqs.~\er{Pi} and \er{Pi_til}), the diagonal components of $\Pi_{\alpha\beta}$ are all equal at $\Delta_Z=0$, $q\to0$. This reflects the additional contribution of intra-band terms to $\Pi_{xx,yy}$ in the static limit discussed above.

\subsection{Phonon soft mode}
\label{qneq0}
Our goal now is to identify the critical mode of the transition in the static $\Omega=0$ limit. To do that, we study when the phonon eigenfrequencies, renormalized by static self-energy corrections \er{Pi_q<Z}, turn to zero. As discussed in Sec.~\ref{distinction}, for static case the LO-TO splitting is always significant \er{diff2} and must be taken into account. We use the same form of the phonon propagator as in Eq.~\er{bareTO}, but with $\Pi_{\alpha\beta}$ obtained in Eq.~\er{Pi_q<Z}. After substituting Eq.~\er{diff2} into Eq.~\er{bareTO}, we get
\beq
\label{gr_t}
\begin{split}
\mathcal{D}^{-1}_{\alpha\beta} = -\frac{2\pi V}{\Omega_0^2} \Big[&\big( \Omega_m^2 + \omega_\text{TO}^2(\bq) \big) \delta_{\alpha\beta} + \omega_{pi}^2 \frac{q_\alpha q_\beta}{q^2 + \kappa^2} + \tilde{\Pi}_{\alpha\beta}(\bq) \Big],
\end{split}
\eeq
where $\tilde{\Pi}_{\alpha\beta}(\bq) \equiv \Pi_{\alpha\beta}(\bq)/(2\pi V/\Omega_0^2)$.
The roots of $\text{Det}[\hat{\mathcal{D}}^{-1}] = 0$ are collective modes of the system.

We start with discussing the case $q\to0$ (below we will justify that the soft mode is at $q=0$). We find three eigenergies:
 \bse
 \label{mode_q0}
 \beq
 \begin{split}
 \label{w1}
 \Omega_1^2 (q\to0) =& \omega_\text{TO}^2 \bigg[ 1 - \frac{\alpha}{\alpha_0} \bigg\{ 1 - \frac{\Delta_Z^2}{8 (2\mu_0)^2} \bigg\} \bigg], \\
 \end{split}
 \eeq
 \beq
 \label{w2}
 \begin{split}
 \Omega_2^2 (q\to0) =& \omega_\text{TO}^2 \bigg[ 1 - \frac{\alpha}{\alpha_0} \bigg\{ 1 - \frac{\Delta_Z^2}{8 (2\mu_0)^2} \bigg\} \bigg], \\
 \end{split}
 \eeq
 \beq
 \label{w3}
 \begin{split}
 \Omega_3^2 (q\to0) =& \omega_\text{TO}^2 \bigg[ 1 - \frac{\alpha}{\alpha_0} \bigg\{ 1 - \frac{\Delta_Z^2}{4 (2\mu_0)^2} \bigg\} \bigg]. \\
 \end{split}
 \eeq
 \ese
For $\Delta_Z = 0$ it is evident that all the eigenfrequencies turn to zero simultaneously at $\alpha=\alpha_0$, corresponding to the usual $O(3)$ universality class of the transition. At finite $\Delta_Z$, not all of the frequencies in Eq.~\er{mode_q0} go to zero simultaneously on increasing $\alpha$. In particular, $\Omega_1$ and $\Omega_2$ go unstable first. 
Analyzing the $\bq$-dependent eigenvectors of Eq.~\er{gr_t}, we conclude that $\Omega_1$ and $\Omega_2$ in Eq.~\er{mode_q0} are always the modes that are polarized perpendicular to the magnetic field. Consequently, we find that in a spin-orbit coupled polar metal, the presence of the magnetic field introduces an easy plane anisotropy, pinning the polarization of the critical polar mode perpendicular to magnetic fields.

These results can be understood within the framework of Landau theory for the polar order parameter in the presence of a weak magnetic field ${\bf H}$. The free energy takes the following form (consistent with spherical symmetry above transition):
\beq
\label{fe}
F[\bP, {\bf H} ] \approx [a(T)+\eta {\bf H}^2]\bP^2+ \zeta (\bP\cdot {\bf H})^2 +b \bP^4.
\eeq
From the above \er{fe}, it follows that in a polar metal with spin-orbit coupling, the conduction electrons result in $\eta>0$ and $\zeta>0$. Close to the transition, this implies the possibility to orient the polar order orthogonal to the applied magnetic field.

We now turn to the case of finite $q$. In the $q^2/2m_b \ll v_Fq \ll \Delta_Z$ regime, the $q^2$ corrections to the eigenfrequencies in Eq.~\er{mode_q0} are
\bse
\label{mode_q<Z}
\beq
\begin{split}
\label{wq1}
\Omega_1^2(q) =& \Omega_1^2(q\to0) \\
&+ \bigg[ c_T^2 + \frac{\alpha}{\alpha_0} \frac{\omega_\text{TO}^2}{8k_F^2} \bigg\{ 1 - \frac{17}{160} \frac{\Delta_Z^2}{(2\mu_0)^2} \bigg\} \bigg] (q_x^2+q_y^2) \\
&+ \bigg[ c_T^2 + \frac{\alpha}{\alpha_0} \frac{\omega_\text{TO}^2}{8k_F^2} \bigg\{ 1 - \frac{11}{160} \frac{\Delta_Z^2}{(2\mu_0)^2} \bigg\} \bigg] q_z^2
\end{split}
\eeq
\beq
\label{wq2}
\begin{split}
\Omega_2^2(q) =& \Omega_2^2(q\to0) \\
&+ \bigg[ c_L^2 + \frac{\alpha}{\alpha_0} \frac{\omega_\text{TO}^2}{4k_F^2} \bigg\{ 1 - \frac{57}{320} \frac{\Delta_Z^2}{(2\mu_0)^2} \bigg\} \bigg] (q_x^2+q_y^2) \\
&+ \bigg[ c_T^2 + \frac{\alpha}{\alpha_0} \frac{\omega_\text{TO}^2}{8k_F^2} \bigg\{ 1 - \frac{11}{160} \frac{\Delta_Z^2}{(2\mu_0)^2} \bigg\} \bigg] q_z^2
\end{split}
\eeq
\beq
\label{wq3}
\begin{split}
\Omega_3^2(q) =& \Omega_3^2(q\to0) \\
&+ \bigg[ c_T^2 + \frac{\alpha}{\alpha_0} \frac{\omega_\text{TO}^2}{8k_F^2} \bigg\{ 1 + \frac{3}{40} \frac{\Delta_Z^2}{(2\mu_0)^2} \bigg\} \bigg] (q_x^2+q_y^2) \\
&+ \bigg[ c_L^2 + \frac{\alpha}{\alpha_0} \frac{\omega_\text{TO}^2}{4k_F^2} \bigg\{ 1 - \frac{17}{160} \frac{\Delta_Z^2}{(2\mu_0)^2} \bigg\} \bigg] q_z^2
\end{split}
\eeq
\ese
where $\Omega_i^2(q\to0)$ is given in Eq.~\er{mode_q0}, $c_{T(L)}$ is the TO(LO) phonon velocity with $c_L^2 = c_T^2 + \omega_{pi}^2/(q^2 + \kappa^2)$ and, $\alpha$ and $\alpha_0$ are defined in Eqs.~\er{alpha} and \er{a0}, respectively.

One observes that the rotational symmetry in the $xy$-plane (the plane perpendicular to the direction of magnetic field) results in the degeneracy of $\Omega_1$ and $\Omega_2$ at $q\to0$ \er{mode_q0}, which is broken by finite $q$ in the plane \er{mode_q<Z}. 
Moreover, for $\Delta_Z\ll \mu_0$ the momentum dispersion of each mode in Eq.~\er{mode_q<Z} (terms of order $\sim q^2$), remains positive. This result also holds in the opposite regime when $\Delta_Z \ll q^2/2m_b \ll v_Fq$; see Appendix~\ref{appen:q} for technical details. Consequently, the zero-energy solution first appears at $q=0$, proving that the critical mode of the polar transition remains at $q=0$ when coupled to conduction electrons.

\section{Discussion}
\label{discuss}
In this section, we extend the results of Secs.~\ref{zero_q} and \ref{finite_q} and discuss their application to experiments. We first address the effects of temperature, orbital magnetic field effects and finite lifetimes on our results. Next we present extension of the results of Sec.~\ref{zero_q} to the  
2D case, given the recent discovery of several  2D polar metals\cite{Fei:2018,Cao:2018}. We conclude this section considering the experimental feasibility of extracting the coupling constant $\lambda$ in both 3D and 2D,
and also discuss the field-orientation of the polar order.

\subsection{Applicability of approximations}
\label{approx}
For most of our derivations in Sec.~\ref{model} and results in Secs.~\ref{zero_q} and \ref{finite_q} we assumed a $T=0$, clean system, and we ignored orbital quantization in magnetic field. In this section we will discuss the effects of relaxing these approximations and assess their applicability.

\subsubsection{Finite temperature effects}
\label{sec:finiteT}
Finite temperature effects can be analyzed using the general expression for the phonon self-energy, Eq. \eqref{se1}. One observes that $T$ enters the expression only via the Fermi functions. The $k$-integral for all the components of the phonon self-energy \er{se1} has the same general form which can be written after variable substitution as
\beq
\label{zT}
\begin{split}
\Pi_{\alpha\beta}(\Omega_m) \sim \sum_{s \bar{s}} g_{\alpha\beta}^{s \bar{s}}(\Omega_m) \int_{-\mu}^\infty d\xi (\xi&+\mu)^{3/2} \Big[ n_F(\xi - \Delta_Z/2) \\
- & n_F(\xi + \Delta_Z/2) \Big],
\end{split}
\eeq
where $\xi = \ve_\bk - \mu$ and $g_{\alpha\beta}^{s \bar{s}}(\Omega_m) \sim \big( i\Omega_m - (s-\bar{s})\Delta_Z/2 \big)^{-1} $. Assuming temperatures $T \ll \mu$, the lower limit of $\xi$-integration in Eq.~\er{zT} can be extended to $-\infty$. 
Now we consider two regimes: $\Delta_Z \ll T$ and $\Delta_Z \gg T$. In the $\Delta_Z \ll T$ regime, one can expand the difference of two Fermi functions in Eq.~\er{zT}, yielding a prefactor proportional to $\Delta_Z/T$. However, after performing the integral to the leading order in $T/\mu$ all the $T$-dependence drops out: the $\xi$-integral gives $T \mu^{3/2} (\Delta_Z/T) \sim \mu^{3/2}\Delta_Z$. Likewise, in the opposite regime, i.e., $\Delta_Z \gg T$,one finds that the integral is $\propto \Delta_Z$. Indeed, the explicit calculation suggests that the $T$-dependent correction appears as $(T/\mu)^2$ in both limits. The final form of $\Pi_{ij}$ in both the regimes, therefore, reads
\beq
\label{zT_f}
\begin{split}
\Pi_{ij}(\Omega_m) \propto \Delta_Z \bigg\{1 + \mathcal{O} \Big( \frac{T}{\mu} \Big)^2 \bigg\} + \frac{\Delta_Z^3}{32 \mu^2} \bigg\{1 + \mathcal{O} \Big( \frac{T}{\mu} \Big)^2 \bigg\} +...
\end{split}
\eeq
So the conclusion is that in both the regimes, the $T$-dependent correction is small, i.e., $\mathcal{O}(T/\mu)^2$, regardless of the $\Delta_Z$ value.

\subsubsection{Resonance broadening due to scattering}
\label{disorder}
We next discuss the broadening of $q=0$ eigenmodes due to scattering of electrons by non-magnetic disorder or interactions. To make our point, for simplicity, we consider the high carrier density regime when there is no distinction between LO and TO mode frequencies. We assume the broadening effect to be characterized by the scattering rate $\Gamma$ which affects only the electron Green's function; phonons are assumed to be sharp in our treatment. 

In the phonon self-energy, the presence of disorder simply amounts to replacing $i\Omega_m$ everywhere in Eq.~\er{Pi} by $\Omega + i\Gamma$ and obtain real and imaginary parts; see Appendix~\ref{app:damping} for details. The imaginary part shows spin-flip resonances peaked at $\Omega = \Delta_Z$ with $\Gamma$ as the width of the resonance due to disorder. 

We now analyze the effect of this broadening on the hybridized collective modes arising from the interaction \eqref{coupling}. 
We illustrate the effect for the energies near the anticrossing point ($\Delta_Z \approx \omega_\text{TO}$) and at high carrier densities, relevant for the discussion in Sec.~\ref{exp}. 
We then need to consider the modes $\Omega_2$ and $\Omega_5$, given in Eq.~\er{w25} and shown in Figs.~\ref{q=0}(a) and (b), which give maximum splitting near the anticrossing point. Assuming $\Gamma$ to be small compared to the real part of the mode frequencies one obtains
\beq
\label{scatt}
\begin{split}
\Omega_2^\Gamma &= \Omega_2 - \frac{i \Gamma}{2} \Bigg[ 1 - \frac{\omega_\text{TO}^2 - \Delta_Z^2}{\sqrt{\frac{8\alpha}{15} \Delta_Z L_Z + (\omega_\text{TO}^2 - \Delta_Z^2)^2}} \Bigg], \\
\Omega_5^\Gamma &= \Omega_5 - \frac{i \Gamma}{2} \Bigg[ 1 + \frac{\omega_\text{TO}^2 - \Delta_Z^2}{\sqrt{\frac{8\alpha}{15} \Delta_Z L_Z + (\omega_\text{TO}^2 - \Delta_Z^2)^2}} \Bigg],
\end{split}
\eeq
where $\Omega_{2,5}$ is given in Eq.~\er{w25}. The Eq.~\er{scatt} can be read as resonance peaks at $\Omega_2^\Gamma$ and $\Omega_5^\Gamma$ with widths given by the imaginary part. As we can see that near the anticrossing ($\omega_\text{TO} \approx \Delta_Z$), the second term of the imaginary part vanishes. Therefore, the effect of disorder leads to the width of resonance for both the modes to be $\Gamma/2$. 

Finally, in between $\Omega_2$ and $\Omega_5$ in Figs.~\ref{q=0}(a) and (b), there exists another mode $\Omega_3$ which also gets broadened due to disorder. If we assume that the maximum width of $\Omega_3$ is $\Gamma$, then to resolve all three modes the energy splitting must be larger than $3\Gamma/2$.

\subsubsection{Orbital effects}
In our discussion earlier, throughout we only considered Zeeman effects of the magnetic field and ignored its orbital effects. The orbital effect of the magnetic field implies Landau quantization of the electronic levels.\cite{landau}
In this section we will argue that finite $T$ and scattering effects smear out orbital effects allowing to neglect them with respect to the Zeeman splitting.

Physically, the Landau quantization leads to oscillations of various physical quantities, such as density of states or resistivity in the presence of magnetic fields. Finite temperatures reduce the oscillation amplitude by the factor $R_T R_\sigma$\cite{shoenberg:1984} given by
\beq
\label{red_fact}
R_T = \frac{2\pi^2 p k_B T/\hbar \omega_c}{\sinh(2\pi^2 p k_B T/\hbar \omega_c)}; \,\,\, R_\sigma = \cos \Big( \pi p \frac{\Delta_Z}{\hbar \omega_c} \Big)
\eeq
where $k_B$ is the Boltzmann constant, $\omega_c = eB/m_c c$ is the cyclotron frequency with $m_c$ as the cyclotron effective mass, and $p$ is the number of the oscillation harmonic. Here $R_\sigma$ comes from the energy difference $\hbar\omega_c$ between spin-up and spin-down electrons in the magnetic field. The ratio $\Delta_Z/\hbar\omega_c$ can be written as $gm_c/2m_e$, where $m_e$ is the free electron mass and $m_c$ is defined as $m_c = (\partial A_e/\partial \ve)/2\pi$ with $A_e$ as the extremal cross-sectional area. In a free-electron system, $g \approx 2$ and $A_e = \pi k_F^2 = 2\pi m_e\mu_0$, which means $m_c=m_e$. The Zeeman term thus becomes $(-1)^p$.

If the magnetic field, cyclotron effective mass and temperatures are such that $k_B T \gg \hbar \omega_c$, we have 
\beq
\label{red1}
R_T = \frac{4\pi^2 p k_B T}{\hbar \omega_c} \text{exp} \Big( -\frac{2\pi^2 p k_B T}{\hbar \omega_c} \Big).
\eeq
For practical conditions of $T$ and $B$, and $m_c$ not very different from $m_e$, the above Eq.~\er{red1} is a good approximation. Assuming that the electron mass is not very different from the cyclotron effective mass and also the $g$-factor is of order $1$, one has $\hbar \omega_c \sim \Delta_Z$. In this case, for $T\gg\Delta_Z$, the oscillation amplitude is suppressed exponentially, resulting in the smearing of Landau levels, while the spin-phonon resonances, Figs.~\ref{q=0} and \ref{split_q=0}, are still expected to be observed. As has been shown above,  for $T \gg \Delta_Z$, the leading finite $T$ correction to the self-energy is of order $(T/\mu_0)^2$, which is small in our assumption. The spin-phonon resonances, therefore, are not significantly affected. Thus, at elevated temperatures, the orbital effects can be ignored. 

In the opposite regime, i.e., $k_B T \ll \hbar \omega_c \sim \Delta_Z$, the coupled spin-phonon modes and Landau levels both are sharp and on this ground the orbital effect ($R_T = 1$) seems to play a significant role too. This is where scattering effects play an important role and we will argue that the orbital effects are suppressed again.

Finite electronic scattering rate (due to disorder or interactions) suppresses both orbital and spin effects of the magnetic field. If the electrons have a finite relaxation time $(\tau)$, parametrized by the damping factor $\Gamma$ $(\tau \sim \Gamma^{-1})$, the otherwise sharp Landau levels broaden leading to reduction of the oscillation amplitude.\cite{dingle:b} This implies an additional reduction factor, known as the Dingle reduction factor $(R_D)$,\cite{dingle:b, shoenberg:1984}
\beq
R_D = \text{exp} \Big( -\frac{2\pi p \Gamma}{\hbar \omega_c} \Big),
\eeq 
that is appended with the oscillation amplitude.
We see that the additional factor $R_D$ (the overall oscillation amplitude becomes $R_T R_\sigma R_D$ in the presence of disorder) suppresses the oscillation amplitude exponentially. The spin-phonon resonances are also smeared due to disorder, as discussed in Sec.~\ref{disorder}, however, the exponential suppression of the orbital effects can be much stronger at low to moderate fields. This suggests that for real not too clean systems, the orbital quantization effects may be negligible also at low temperatures. We note, however, that in the absence of disorder, a mixing between the orbital cyclotron resonance and optical phonons has been theoretically predicted \cite{Abergel08} for 2D electrons in quantum wells.

\subsection{Summary of 2D systems}
\label{2D}
Systems such as few layers of WTe$_2$\cite{Fei:2018} and tri-layer superlattices\cite{Cao:2018} have been reported to be 2D polar metals. With the advent of these materials, it is therefore imperative to study the interaction of electronic excitations and phonons in 2D as well. For the purpose of this section we assume $\Omega \gg v_Fq$ where it is legitimate to consider $q=0$ as discussed in Sec.~\ref{zero_q}. Unlike the case in 3D, at $q=0$ the LO-TO distinction vanishes for 2D, and there is no need to go to the high carrier density regime.\cite{Cudazzo:2011,Sohier:2017}

In particular, we consider the applied magnetic field to be in the $xy$ plane. We assume an isotropic 2D electron dispersion and choose the field to be along $x$ axis. Upon explicit calculation at $T=0$, the self-energy matrix reads:
\beq
\label{Pi_2D}
\hat{\Pi}(\Omega_m) = - \frac{\lambda^2 m_b^2 A \mu_0}{\pi} \frac{\Delta_Z}{\Omega_m^2 + \Delta_Z^2} \hat{\tilde{\Pi}}_{2D}
\eeq
where
\beq
\label{pi_mat}
\hat{\tilde{\Pi}}_{2D} =
\begin{pmatrix}
\Delta_Z & 0 & 0 \\
0 & \Delta_Z & -\Omega_m \\
0 & \Omega_m & \Delta_Z
\end{pmatrix}.
\eeq
and $A$ is area of the crystal. Note that the chemical potential in 2D is unaffected by magnetic fields; see Sec.~\ref{chem_2D}.

Physically, the self-energy matrix \er{Pi_2D} and \er{pi_mat} can be understood directly from its 3D counterpart as discussed in Sec.~\ref{SE_q=0}. Since the field is applied along $x$-direction and for 2D $k_z=0$, one can immediately say that for $\Pi_{xx}$ only $j_y^{\sigma_z} - j_y^{\sigma_z}$ correlation contributes. Similarly for $\Pi_{yy}$, the only contribution is from $j_y^{\sigma_z} - j_y^{\sigma_z}$ correlation. $\Pi_{zz}$ consists of four terms: $j_x^{\sigma_y} - j_x^{\sigma_y}$, $j_y^{\sigma_x} - j_x^{\sigma_y}$, $j_x^{\sigma_y} - j_y^{\sigma_x}$ and $j_y^{\sigma_x} - j_y^{\sigma_x}$ correlations, of which only $j_x^{\sigma_y} - j_x^{\sigma_y}$ contributes. So, unlike for 3D, in 2D one anticipates $\Pi_{xx} = \Pi_{yy} = \Pi_{zz}$ already by symmetry. Using similar arguments for the off-diagonal contributions one finds that since field is applied along $x$-direction, only $\Pi_{yz}$ and $\Pi_{zy}$ remain nonzero \er{pi_mat}.

Finally, we calculate the coupling of modes the same way as that for 3D in Sec.~\ref{hybrid_B}. We note that in general, the phonon polarized along $z$ should not have the same energy, as the ones polarized in $xy$ plane, as they belong to different irreducible representations. Here we neglect this splitting to illustrate the qualitative effects; it can be included into consideration straightforwardly by modifying the phonon propagator with an additional term $\sim \delta_{\alpha,3}\delta_{\beta,3}$. The 2D analogue of Eq.~\er{full}, but without LO-TO splitting, can be written as
\beq
\label{full_2D}
\begin{split}
\text{Det} \bigg[ \big( \Omega_m^2 + \omega_\text{TO}^2 \big) \hat{\mathbb{1}} - \alpha_{2D} \frac{\Delta_Z}{\Omega_m^2 + \Delta_Z^2} \hat{\tilde{\Pi}}_{2D} \bigg] = 0,
\end{split}
\eeq
where $\hat{\tilde{\Pi}}_{2D}$ is given in Eq.~\er{pi_mat} and the coupling constant $\lambda$ is parametrized by
\beq
\label{a2D}
\alpha_{2D} \equiv \frac{\lambda^2 m_b^2 \Omega_0^2 A \mu_0}{2\pi^2 V},
\eeq
having the unit of $Energy^2$. Note that $\alpha_{2D}$ \er{a2D} consists of the ratio $A/V$, which in experimental terms is equivalent to $1/a_0$, where $a_0$ is the lattice constant. Solving Eq.~\er{full_2D} for $\Omega$, we get five positive definite roots, regarded as spin-phonon resonances, which disperse with $\Delta_Z$. In this case also only the lowest lying mode (2D analogue of $\Omega_5$ in Eq.~\er{small_Z} and in Fig.~\ref{q=0}) that becomes unstable at weak fields for $\alpha_{2D} > \omega_\text{TO}^2$.

Eigenfrequencies that give maximum splitting near the anticrossing point reads:
\beq
\Omega_{2(5)} = \frac{1}{\sqrt{2}} \bigg[ \omega_\text{TO}^2 + \Delta_Z^2 \pm \sqrt{(\omega_\text{TO}^2 - \Delta_Z^2)^2 + 4\alpha_{2D} \Delta_Z^2} \bigg]^{1/2},
\eeq
where $\Omega_2$ is the one with `+' sign, while $\Omega_5$ is the one with `-' sign. Following the same procedure as discussed for the 3D case in Sec.~\ref{ws<wp}, we obtain the coupling constant:
\beq
\label{lambda_2D}
\lambda = \sqrt{\frac{\pi a_0 (\Delta\Omega^2)^2}{2 n m_b \Delta_Z^2 \Omega_0^2}}.
\eeq
Similar to the 3D case, from the measured $\Delta\Omega^2$ at a particular carrier density and magnetic field, one can determine $\lambda$ for known $m_b$, $a_0$ and $\Omega_0$ of the material.

\subsection{Experimental Proposal}
\label{exp}
In this section, we will discuss the prospect to measure the electron-phonon coupling constant $\lambda$ \er{coupling} and to orient the polar order by magnetic field. We begin with the former, assuming the case of high density metals when LO-TO distinction can be ignored. In Fig.~\ref{q=0}, it has been shown that the excitations of the nearly polar metal exhibit an anticrossing at $\Delta_Z\approx\omega_\text{TO}$.
The magnitude of this anticrossing allows to extract the value of $\lambda$ directly (Eq. \eqref{lambda}). Note that it follows from the definition of Green's function, Eqs.~\er{full} and \er{alpha}, that the true expansion parameter for the electron-phonon interaction is $\lambda \Omega_0$. Consequently, while the ionic frequency $\Omega_0 = \sqrt{\varepsilon_\infty} \omega_{pi}$ may not be straightforward to extract from measurements (especially in the strongly metallic case), it is sufficient to deduce for the 3D case
\begin{equation}
   \lambda_\text{phys} \equiv \lambda \Omega_0= \sqrt{\frac{\pi (\Delta\Omega^2)^2}{4 n m_b\Delta_Z^2 }},
   \label{eq:lambdaphys}
\end{equation}
where $\Delta\Omega^2$ is the difference between frequencies squared (see Eq. \er{diff}). All quantities on the RHS of Eq. \eqref{eq:lambdaphys} are experimentally measurable. To estimate the value of $\Delta\Omega$, which could be relevant in determining possible experimental probes, we cast $\lambda\Omega_0$ in Eq.~\er{eq:lambdaphys} in terms of the parameters used in Refs.~\onlinecite{gastiasoro2021theory} and \onlinecite{Ruhman:2016} for the interaction Hamiltonian. Recalling that $\Omega_0 \equiv \sqrt{\ve_\infty} \omega_{pi}$, we can write $\lambda\Omega_0 \equiv \delta t/\sqrt{\rho}$, where $\delta t \equiv n_0 (Ze) \lambda$ is the electron-phonon coupling constant parametrized in Ref.~\onlinecite{Ruhman:2016} and $\rho \equiv Mn_0$ is the ionic mass density with $M$ and $n_0$ as reduced ionic mass and ionic density, respectively. Replacing $\Delta_Z$ by $\omega_\text{TO}$ (valid near the anticrossing point), we get for $\Delta\Omega$ as (assuming weak splitting $\Delta \Omega^2\approx 2 \omega_\text{TO} \Delta \Omega $)
\beq
\label{estimate}
\Delta\Omega \approx \sqrt{ \frac{nm_b \delta t^2}{\pi \rho}}.
\eeq


The value of the splitting of the collective mode frequencies, $\Delta\Omega$ \er{estimate}, can be measured in spectroscopic experiments that can detect $q=0$ phonons. The most promising way to study the phonon excitations is undoubtedly inelastic neutron scattering; in particular, powder experiments are sufficient, since the splitting can be observed at $q=0$. For low-density metals or doped semiconductors (such that the light can penetrate sufficiently deep in the sample), IR spectroscopy can be alternatively used. Both techniques are primarily sensitive to the phonon excitations, meaning that the spin-flip mode itself would be harder to observe. We note that techniques that couple to spin susceptibility (e.g., ESR), will not observe the modes described here; as has been mentioned above, the spin-flip modes we consider here arise from the spin-current response function, and not spin susceptibility.

The important limiting factor in real experiments would be the magnetic field range needed for the experiments. Assuming the $g$-factor of conduction electrons to have the free-electron value, the required field is of the order of $10$ T per $1$ meV of the transverse phonon energy $\omega_\text{TO}$. We note that for a second order transition, $\omega_\text{TO}$ would go to zero at the critical temperature (or value of external parameter for a quantum transition), and thus the required field can be minimized by going close to the transition. To give an example, for SrTiO$_3$ the relevant phonon energy is around 2 meV at low temperature \cite{yamada}, suggesting a field of around 20 T (see also below), accessible in modern laboratories. Using Ca-substitution \cite{Rischau2017, Engelmayer2019b, Wang2019}, however,  SrTiO$_3$ can be driven ferroelectric, allowing for measurements at lower fields due to softening of the TO phonon.

While the temperature, as discussed above, does not affect the observation of the anticrossing, the electronic scattering rate does. In usual metals, the scattering times are of the order $10^{-14}$ s \cite{scatteringtime1,scatteringtime2} at room temperature, implying $\Gamma\sim 60$ meV and fields in excess of 600 T to observe the splitting. On the other hand, as the typical residual resistivity ratios can be often of the order 100, this implies that disorder scattering in clean samples at low temperatures restricts the field to being above 6 T or less only. This order-of-magnitude estimates suggest that for polar metals with classical transitions well below room temperature the spin-phonon resonance can be potentially observed at fields below 10 T.

We now turn to the possibility of orienting the polar order with magnetic field, discussed in Sec. \ref{finite_q}. In particular, we found that the polar order is "pinned" to the plane orthogonal to the field direction (see Fig. \ref{fig:intro}). The argument above can be readily generalized to systems of lower symmetry. For example, if we consider the tetragonal symmetry, as considered in SrTiO$_3$,\cite{bednorz1984} the polar order will be along the pseudocubic [110] or [1$\overline{1}$0] axis without magnetic field. For doped SrTiO$_3$, we then predict that orienting the magnetic field along one of these axes would pin the polar order along the other axis, allowing on-demand switching of polar order by magnetic fields. This presents an attractive 
alternative\cite{Fei:2018,Sharma19,Zabala21} to electric-field switching that is hard to apply in bulk materials due to screening by conduction electrons. The resulting switching can be probed by transport anisotropy. Indeed, the polar order breaks rotational symmetry, as happens also in nematic phases, where resistivity anisotropy is the conventional probe to detect this state \cite{fradkin2010}. Field orientation of the polar order would then result in a change in transport anisotropy, and could have applications in spintronics.

We next discuss the candidate materials where the effects discussed here could be observed. Recently, LiOsO$_3$\cite{shi:2013, laurita:2019, jin:2019} and few-layer WTe$_2$\cite{Fei:2018} have been observed to be intrinsic 3D and 2D polar metals, respectively. On the other hand, there has been also evidences of artificial 2D metals\cite{Cao:2018}, and 3D doped systems such as SrTiO$_3$,\cite{Rischau2017, Wang2019, Engelmayer2019b} BaTiO$_3$,\cite{takahashi:2017, kolodiazhnyi:2010}, and KTaO$_3$\cite{calvi:1995} undergoing polar phase transition. On general grounds one seeks for materials with strong spin-orbit coupling and not too low density: the actual strength of electron-phonon coupling effects, Eq. \eqref{coupling}, grows with both $\lambda$ and electron density (due to $k\sim k_F$).

As far as the perovskites are concerned, in SrTiO$_3$ (STO), $\lambda$ has been proposed to be origin of the superconducting pairing\cite{Ruhman:2016, Maria:2020, Maria:review} and there is recent theoretical suggestion that it can be indeed sizable \cite{gastiasoro2021theory}.
However, STO consists of Ti atoms which are quite light and results in a relatively weak (20 - 30 meV) SOC effect in the band structure.\cite{marel} BaTiO$_3$ can be then expected to have a comparable $\lambda$. Perovskites consisting of heavier elements, such as LiOsO$_3$ or KTaO$_3$ (KTO), are expected to have stronger SOC, and hence stronger $\lambda$. Indeed, for KTO the SOC-related splitting in the band structure of around 400 meV \cite{nakamura:2009, bruno:2019}, an order of magnitude larger than in STO.  For LiOsO$_3$ it has been reported that, unlike other 5$d$ transition-metal oxides, the effect of SOC on the band structure is small \cite{liu:2015}. Nonetheless, the presence of heavy elements in this polar metal may lead to a large value of $\lambda$ which should be checked experimentally.

Let us now discuss in more detail the most well-studied candidates, STO and KTO.
In the insulating phase, the values of $\omega_\text{TO}$ for STO is known from various experimental means such as inelastic neutron scattering,\cite{yamada} Raman spectroscopy\cite{worlock,fleuryphys} and hyper-Raman spectroscopy,\cite{vogt1995,hehlen_1999,yamanaka_2000} while that for KTO is known from inelastic neutron scattering\cite{shirane:1967}, Infrared\cite{perry:1969} and Raman spectroscopy\cite{fleury,fleuryphys}. These numbers are roughly 1.9 meV\cite{yamada} and 3.1 meV\cite{shirane:1967} for STO and KTO, respectively. Without any modifications moving the material closer to polar quantum critical point, we estimate the required magnetic field near the anticrossing point, $B \approx \omega_\text{TO}/g\mu_B$, for $g$-factors 1.978\cite{Glinchuk2001} for STO and 2.31\cite{golovina:2012} for KTO, to be around 17 T and 23 T, respectively. The energy scale $\Omega_0$ also is an experimentally accessible parameter and for STO and KTO in their insulating phases is known from inelastic neutron measurements to be 194.4 meV\cite{yamada} and 168 meV\cite{shirane:1967}, respectively. Using these values, the required carrier density to ignore LO-TO splitting (such that $\omega_{p\infty} \gg \omega_{pi}$), comes out to be around $10^{19}-10^{20} \text{cm}^{-3}$. 

Now we provide an estimate for the mode splitting, $\Delta\Omega$ \er{estimate}, in strongly doped STO near the anticrossing point. The DFT calculated value of $\delta t$ that appears in Eq.~\er{estimate} for the case of STO suggests $\delta t = t_{xy}' a_0 \approx 421$ meV \cite{gastiasoro2021theory}, where $a_0 = 3.9 \AA$ is the lattice constant for STO. However, the pairing interaction energy for this value of $\delta t$ is too weak for superconducting $T_c$ to be consistent with experiments. To get the values of $T_c$ in the relevant range, the BCS coupling constant $\lambda_\text{BCS} = \nu_F V_\text{eff}$, where $\nu_F$ is electronic density of states and $V_\text{eff}$ is effective attractive interaction which implicitly depends on $\delta t$, needs to be cranked up. This can be achieved by going near the polar quantum critical point (QCP), as shown in Ref.~\onlinecite{gastiasoro2021theory}. Using the values obtained in Ref.~\onlinecite{gastiasoro2021theory}, we find the cranked up $V_\text{eff}$ to be around 2 eV, which gives new $\delta t$ to be roughly 5 times the DFT value. For this new $\delta t$, and using the experimentally known parameters of STO, i.e., $m_b = 7m_e$\cite{ARPESSTO2010} with $m_e$ as bare electronic mass, and $\rho \approx 5 \text{g}/\text{cm}^3$\cite{Ruhman:2016}, we obtain the mode splitting $\Delta\Omega$ \er{estimate} to be around 0.4 meV at carrier densities of order $10^{20} \text{cm}^{-3}$. This is the required value of $\delta\Omega$ at the given carrier density if superconductivity in STO is indeed driven by Rashba-type electron-phonon coupling \er{coupling}. For even higher carrier density, for e.g., $10^{21} \text{cm}^{-3}$, $\Delta\Omega \approx 1.3$ meV (assuming same effective mass is applicable).
We note that in doped STO, the TO phonon frequency also increases with density: using the relation in Ref.~\onlinecite{Bauerle1980} with a value of 1.9 meV\cite{yamada} for the undoped case, we get $\omega_\text{TO}\approx 4.6$ and $13.5$ meV for $n=10^{20}\text{cm}^{-3}$ and $n=10^{21}\text{cm}^{-3}$, respectively. This corresponds to magnetic fields of around 50 and 140 T. These numbers, however, can be reduced by tuning the system closer to the polar QCP: isovalent substitution or strain can be used for that matter \cite{Burke1971,Uwe1976,Wang2019}. In particular, for system sufficiently close to QCP, $\omega_\text{TO}$ can be made arbitrarily small, reducing the required fields to be in the practically attainable range.

Doped KTO may also be a promising candidate for the observation of the effects discussed here: with the spin-orbit coupling effects being roughly 20 times larger than that in STO,\cite{nakamura:2009,bruno:2019} one can expect the splitting $\Delta \Omega$ to be (using $\delta t\approx 20\delta t_\text{STO}$, with $\delta t_\text{STO} \approx 421$ meV \cite{gastiasoro2021theory}, $m_b\approx 0.8 m_e$\cite{KTOmass}) around 0.6 meV for $n=10^{20}\text{cm}^{-3}$ (which is achievable with chemical doping\cite{Sakai_2009}).

From the above discussion, it is clear, however, that an ideal system for the observation of effects discussed here would be an intrinsic polar metal with large density of itinerant electrons (due to the density dependence of \er{estimate}); LiOsO$_3$ may provide such an example.

\section{Conclusions}
\label{conclu}
In this paper, we have analyzed interactions between the collective modes and the order parameter fluctuations of metallic systems close to polar transitions in the presence of magnetic fields. More specifically we have considered the effects of spin-orbit 
coupling mediated interactions between electrons and phonons, both transverse and longitudinal, as allowed by symmetry. 
We have found that phonons, plasmons and magnetic field-induced spin-flip transitions all can hybridize leading to anticrossings in the spectra (Figs. \ref{q=0} and \ref{split_q=0}). We have demonstrated that the splitting energies at the anticrossings can be used to determine the strength of the spin-orbit coupling mediated interactions between electrons and phonons in spectroscopic experiments, such as inelastic neutron scattering or IR spectroscopy (Sec. \ref{exp}). The polar transition manifests itself in the softening of the effective $g$-factor of a hybridized spin-flip excitation at the transition.  Additionally we have shown (Sec. \ref{finite_q}) that the order parameter of the polar phase can be oriented by magnetic field in a polar metal; in particular, the electron-phonon interaction pins it to the plane orthogonal to the field direction.
The proposed measurements will put important constraints on theoretical descriptions of polar metals, particularly in their
superconducting states where there are many mysteries.  
More generally our results facilitate the search for new types of superconducting and non-Fermi liquid states near polar critical points and suggest new ways to control polar order in metallic systems that holds promise for future applications.

\section{Acknowledgement}
We thank G. Aeppli, P. Armitage, P. Coleman, A. Klein, D. Maslov, E. Shimshoni and Y. Yanase for discussions.  
A. K.\ and P. C.\ are 
supported by DOE Basic Energy Sciences grant DE-SC0020353 and 
P.A.V.\ is supported by a Rutgers Center for Material Theory Postdoctoral Fellowship.
P.A.V.\ and P. C. \ acknowledge the Aspen Center for Physics where part of this work was performed, which is supported by National Science Foundation grant PHY-1607611. This work was partially supported by a grant from the Simons Foundation (P.A.V.).

\appendix
\section{Effect of magnetic field on the chemical potential} \label{chem_B}
For the purpose of our problem, we do not constrain the largeness of applied field and in general it can be such that the system is either partially polarized, when both the spin-split conduction bands are partially occupied (two-band case), or fully polarized, when only the lower conduction band is partially occupied (one-band case). The former corresponds to the case when chemical potential is the largest energy scale in the system while the latter to when Zeeman energy is the largest scale. In this section, we briefly review the effect of magnetic field on the chemical potential in both three-dimension (3D) and two-dimension (2D).

\subsection{Three dimensions}
\label{app:chem_3D}
\begin{figure}
\centering
\includegraphics[scale=0.35]{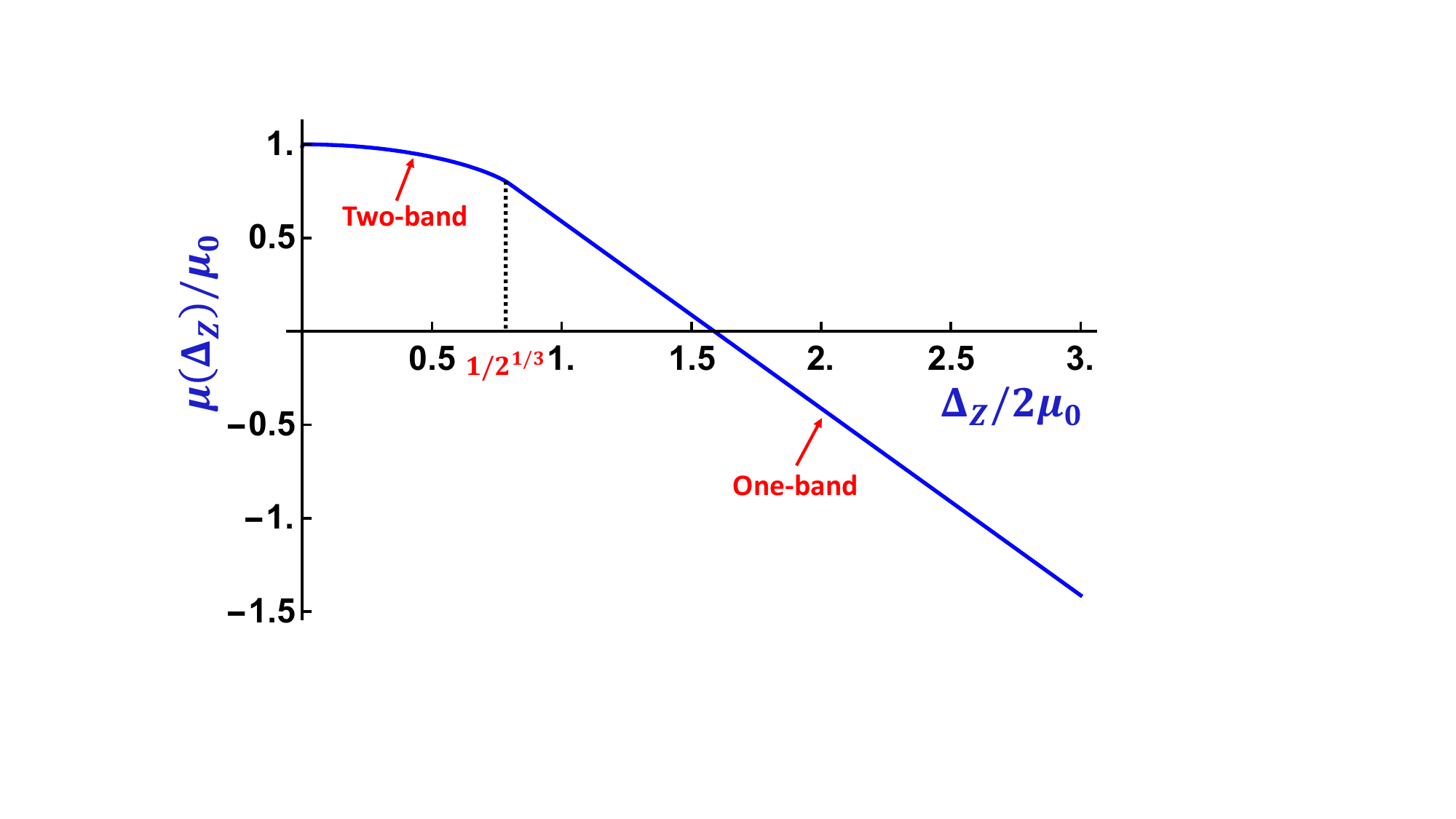}
\caption{\label{chem_pot} Variation of chemical potential with magnetic field}
\end{figure}

The exact form of the chemical potential, obtained numerically, as a function of magnetic field is plotted in Fig.~\ref{chem_pot}. 
One observes that the crossover from two-band regime to one-band regime occurs at $\Delta_Z = 2\mu_0/2^{1/3}$. To understand this result analytically, we provide the derivation for two-band and one-band cases separately. 

\paragraph{Two-band case} 
The spin-up and spin-down densities are $n_\up = [m_b(2\mu(\Delta_Z) - \Delta_Z)]^{3/2}/6\pi^2$ and $n_\down = [m_b(2\mu(\Delta_Z) + \Delta_Z)]^{3/2}/6\pi^2$. The total number density is the sum of two:
\beq
n = \frac{m_b^{3/2}}{6\pi^2} \big[ \big( 2\mu(\Delta_Z) - \Delta_Z \big)^{3/2} + \big( 2\mu(\Delta_Z) + \Delta_Z \big)^{3/2} \big],
\eeq
which can also be written as
\beq
\label{cp_sol}
2 \bigg( \frac{2\mu_0}{\Delta_Z} \bigg)^{3/2} = \bigg( \frac{2\mu(\Delta_Z)}{\Delta_Z} - 1 \bigg)^{3/2} + \bigg( \frac{2\mu(\Delta_Z)}{\Delta_Z} + 1 \bigg)^{3/2}.
\eeq
Here $\mu_0 = (3\pi^2 n)^{2/3}/2m_b$ is the chemical potential in the absence of magnetic field. The above Eq.~\er{cp_sol} needs to be solved for $\mu(\Delta_Z)$. The solution exists only when $2\mu_0 \geq 2^{1/3} \Delta_Z$. The exact analytical form is too complicated to provide here, however, it can certainly be written in the asymptotic limit when $2\mu_0 \gg \Delta_Z$ (by definition it is $2\mu_0$, instead of $\mu_0$). The perturbative solution of Eq.~\er{cp_sol} is
\beq
\label{chem_2band}
2\mu(\Delta_Z) = 2\mu_0 - \frac{\Delta_Z^2}{4(2\mu_0)} - \frac{\Delta_Z^4}{16(2\mu_0)^3}.
\eeq
As one can see that the leading order correction is quadratic in Zeeman energy.

\paragraph{One-band case} 
This limit corresponds to $2\mu_0 \ll \Delta_Z$. In this case, only the lower conduction band, i.e., spin-down subband, is partially occupied and the spin-up density is zero. So we have the condition,
\beq
n = n_\down = \frac{1}{6\pi^2} \big[ m_b \big( 2\mu(\Delta_Z) + \Delta_Z \big) \big]^{3/2},
\eeq
which can be solved straightforwardly for $\mu(\Delta_Z)$ to give
\beq
\label{chem_1band}
\mu(\Delta_Z) = 2^{2/3} \mu_0 - \frac{\Delta_Z}{2}.
\eeq
While obtaining \er{chem_1band}, we assumed that $2\mu(\Delta_Z) < \Delta_Z$. Applying this condition in Eq.~\er{chem_1band}, it is trivial to see that $2\mu_0 < 2^{1/3} \Delta_Z$. So the actual crossover from two-band to one-band regime occurs at $2\mu_0/\Delta_Z = 2^{1/3}$.

Combining Eqs.~\er{chem_2band} and \er{chem_1band}, we get
\beq
\label{chem_tot}
\begin{split}
\frac{\mu(\Delta_Z)}{\mu_0} = \Bigg[ 1 &- \frac{1}{4} \bigg( \frac{\Delta_Z}{2\mu_0} \bigg)^2 - \frac{1}{16} \bigg( \frac{\Delta_Z}{2\mu_0} \bigg)^4 \Bigg] \Theta \bigg( \frac{1}{2^{1/3}} - \frac{\Delta_Z}{2\mu_0} \bigg) \\
&+ \bigg( 2^{2/3} - \frac{\Delta_Z}{2\mu_0} \bigg) \Theta \bigg( \frac{\Delta_Z}{2\mu_0} - \frac{1}{2^{1/3}} \bigg).
\end{split}
\eeq
Note that the expression of two-band case is obtained perturbatively, whereas the one-band result is exact. 

\subsection{Two dimensions}
\label{chem_2D}
The spin-up and spin-down densities are $n_\up = [m_b(2\mu(\Delta_Z) - \Delta_Z)]/4\pi$ and $n_\down = [m_b(2\mu(\Delta_Z) + \Delta_Z)]/4\pi$. The total density is the sum of $n_\up$ and $n_\down$, which simply gives $n = n_\up + n_\down = m_b \mu(\Delta_Z)/\pi$. This relation suggests that $\mu(\Delta_Z) = \mu_0$, which means that the chemical potential in 2D remains same as that in the absence of the magnetic field. For one-band case, similar to 3D, we have $\mu(\Delta_Z) = 2\mu_0 - \Delta_Z/2$.

\section{Applicability of the Perturbation Theory}
\label{app:pert}
The calculations we have presented in Secs.~\ref{zero_q} and \ref{finite_q}
  have been performed in the disordered phase with $\alpha < \alpha_0$ (apart
  from a phenomenological discussion of Landau theory in Sec.~\ref{finite_q}).
  We have assumed that $\alpha$ is sufficiently small that we need to consider only leading-order diagrams noting that for $\alpha > \alpha_0$ this may not be the case.  In this Appendix we justify our weak coupling assumption used by estimating the effect of critical fluctuations on the electronic self-energy; this we do by calculating its dimensionless coupling constant $(C)$. 
Qualitatively, we find that $C \approx C_1 \text{log}( m c_T'^2/\omega_\text{TO}'^2 )$, where $C_1$ is some other dimensionless parameter (obtained later in this section from explicit calculation; see Eq.~\er{C1}) composed of coupling constant $\lambda$ and material parameters. Here, $\omega_\text{TO}' = \omega_\text{TO} (1 - \alpha/\alpha_0)^{1/2}$ and $c_T'$ are the renormalized TO phonon mass and velocity, respectively. From the earlier estimate in Ref.~\onlinecite{Ruhman:2016}, $C_1 \ll 1$. Therefore, as long as we are away from the renormalized critical point, the perturbation theory is valid and our assumption is controllable. At the critical point, however, i.e., $\omega_\text{TO}' \rigt 0$, the log(..) diverges and the perturbation theory breaks down. At any rate, we are always away from the critical point.

Now we discuss the calculation in detail. The fermionic self-energy for the interaction between electrons and a single TO phonon can be written as
\beq
 \label{single_se}
 \begin{split}
 \Sigma(\bk, \epsilon_m) = -\lambda^2 T \sum_n \int_\bq \sum_{\mu\nu} & \Gamma_{\mu\nu} (\bk, \bq) \frac{-2\omega_\bq}{\omega_n^2 + \omega_\bq^2} \times \\
 &\times G(\bk+\bq, \epsilon_m+\omega_n),
 \end{split}
 \eeq
 where $\int_\bq$ is a shorthand for $\int d^3q/(2\pi)^3$, $G(\bk, \epsilon_m)$ is the Matsubara analog of the electron's Green's function and
 \beq
 \label{single_vertex}
 \begin{split}
 \Gamma_{\mu\nu} (\bk, \bq) = \text{Tr} \bigg[ \Big( \bk + \frac{\bq}{2} \times \bs \Big)_\mu & e_\mu(\bq) e_\nu(\bq) \Big( \bk + \frac{\bq}{2} \times \bs \Big)_\nu \bigg] \\
& \times \frac{\omega_\bq [\ve_0(\bq) - \ve_\infty]}{4\pi}
 \end{split}
 \eeq
 is the electron-phonon interaction vertex. Upon careful calculation of the interaction vertex \er{single_vertex}, the self-energy \er{single_se} at $T=0$ can be written as
 \beq
 \label{single_se1}
\begin{split}
\Sigma(\bk, \epsilon_m) =& \frac{\lambda^2\Omega_0^2}{\pi} \int \frac{d\omega}{2\pi} \int_\bq \frac{1}{i(\epsilon_m+\omega) - \xi_{\bk+\bq}} \frac{1}{\omega^2+\omega_\bq^2} \times \\
& \times \bigg[ k^2 \big( 1 + \cos^2\theta \big) + 2kq\cos^3\theta \\
&\hspace{0.5cm} + \frac{q^2}{2} \bigg\{ \sin^4\theta \bigg( 1 - \frac{\sin^22\phi}{2} \bigg) + \cos^4\theta \bigg\} \bigg].
\end{split}
\eeq
We assume that the external frequency $\epsilon$ is small compared to all the other energy scales in the system. This allows us to expand the electron Green's function in Eq.~\er{single_se1} up to linear order in $i\epsilon_m$. Upon frequency integral, the above Eq.~\er{single_se1}, after analytic continuation and then derivative with respect to $\epsilon$, can be written as
\beq
\label{single_se2}
\begin{split}
\frac{d\Sigma(k, \epsilon)}{d\epsilon} = -\frac{\lambda^2\Omega_0^2}{\pi} & \int_\bq \frac{1}{2\omega_q} \bigg( \frac{\Theta(\xi_{\bk+\bq})}{(\omega_\bq + \xi_{\bk+\bq})^2} + \frac{\Theta(-\xi_{\bk+\bq})}{(\omega_\bq - \xi_{\bk+\bq})^2} \bigg) \times \\
\times & \bigg[ k^2 \big( 1 + \cos^2\theta \big) + 2kq\cos^3\theta \\
&+ \frac{q^2}{2} \bigg\{ \sin^4\theta \bigg( 1 - \frac{\sin^22\phi}{2} \bigg) + \cos^4\theta \bigg\} \bigg].
\end{split}
\eeq
At $k=k_F$, and assuming $q \ll k_F$, the first term of [...] in Eq.~\er{single_se2} gives dominant contribution. The integral over $\cos\theta$ in Eq.~\er{single_se2} has to be done carefully: knowing that $q \ll k_F$, in the first term the $\Theta$-function constrains the lower limit of $\cos\theta$-integral by $(-q/2k_F)$, while in the second term it constrains the upper limit by $(-q/2k_F)$. Up to a prefactor $k_F^2/(v_Fq)^2$, the $\cos\theta$-integral can be written as
\beq
\begin{split}
\label{x-int}
\int_{-q/2k_F}^1 dx \frac{1+x^2}{(x + \frac{q}{2k_F} + \frac{\omega_\bq}{v_F q})^2} + \int_{-1}^{-q/2k_F} dx \frac{1+x^2}{(x + \frac{q}{2k_F} - \frac{\omega_\bq}{v_F q})^2},
\end{split}
\eeq
where $x \equiv \cos\theta$. For $q \gg \omega_\text{TO}'/c_T'$, 
the TO phonon dispersion can be considered as linear in $q$. This also bounds the lower limit of $q$-integral at $\omega_\text{TO}'/c_T'$. Overall, in the limit $q/2k_F \ll c_T'/v_F \ll 1$, the $x$-integral \er{x-int} simply gives $2v_F/c_T'$. Finally, Eq.~\er{single_se2} can be written as
\beq
\begin{split}
\frac{d\Sigma(k, \epsilon)}{d\epsilon} & \approx -\frac{\lambda^2 \Omega_0^2 k_F^2}{c_T'^2 v_F 4\pi^3} \int_{\omega_\text{TO}'/s}^{2k_Fc_T'/v_F} dq q^2 \frac{1}{q^3} \\
& \approx -\frac{\lambda^2 \Omega_0^2 k_F^2}{c_T'^2 v_F 4\pi^3} \text{log} \bigg( \frac{m c_T^2}{\omega_\text{TO}'} \bigg).
\end{split}
\eeq
The dimensionless coupling constant is, therefore,
\beq
\label{C}
C = \nu_F \frac{\lambda^2 \Omega_0^2}{c_T'^2} \frac{1}{4\pi} \text{log} \bigg( \frac{m c_T'^2}{\omega_\text{TO}'} \bigg) \equiv C_1 \text{log} \bigg( \frac{m c_T'^2}{\omega_\text{TO}'} \bigg),
\eeq
where
\beq
\label{C1}
C_1 = \nu_F \frac{\delta t^2 k_F^2}{\rho (c_T'^2 k_F^2)} \frac{1}{4\pi}
\eeq
with $\nu_F = m_b k_F/\pi^2$ as the total electronic density of states, $\rho$ as the ionic mass density, and $\delta t$ as some energy which parametrizes electron-phonon coupling defined in Sec.~\ref{exp} of the main text.

We can compare the dimensionless coupling constant $C$ with $\alpha/\alpha_0$. From Eqs.~\er{alpha} and \er{a0},
\beq
\label{ratio}
\frac{\alpha}{\alpha_0} = \frac{\lambda^2 \Omega_0^2 m_b^{5/2} (2\mu_0)^{3/2}}{3\pi^3 \omega_\text{TO}^2} \equiv \nu \frac{\delta t^2 k_F^2}{\rho \omega_\text{TO}^2} \frac{1}{3\pi},
\eeq
where we recall that $\omega_\text{TO}$ is the bare TO phonon frequency which is finite in our model. The forms of $C$ \er{C} and $\alpha/\alpha_0$ \er{ratio} are equivalent (up to a log) except that in the latter, by definition, it's the TO phonon mass that appears. This makes sense in the regime when phonon dispersion is such that $\omega_\text{TO}' \ll c_T q \ll \omega_\text{TO}$; the former condition is used for fermion self-energy while the latter is used for phonon self-energy. The condition $\omega_\text{TO}' = \omega_\text{TO} (1 - \alpha/\alpha_0) < \omega_\text{TO}$ is legitimate because electron-phonon coupling shifts the polar critical point to a lower value. 

Most importantly, one observes that for any finite fixed value of $\omega_\text{TO}$, the self-energy effect can be made small for $\alpha \ll \alpha_0$ (i.e., $\lambda$ being small), justifying the perturbative treatment in the main part of the paper.

\section{Investigation of mode instability at finite $q$: $\Delta_Z \ll q^2/2m \ll v_Fq$}
\label{app:q}
We discussed in Sec.~\ref{finite_q} of the main text that in the $q^2/2m_b \ll v_Fq \ll \Delta_Z$ regime, the dispersion of phonon modes never becomes negative \er{mode_q<Z}. To show that this is also the case when $q^2/2m_b, v_Fq$ shoots through $\Delta_Z$, we explore the opposite regime when $\Delta_Z \ll q^2/2m_b \ll v_Fq$. 

We first calculate phonon self-energy at $\Omega=0$ in this regime which up to order $\Delta_Z^2$ and $q^2$ reads:
\beq
\label{Pi_q}
\begin{split}
\Pi_{\alpha\beta} & (\bq) \approx - \frac{2\pi V}{\Omega_0^2}\frac{\alpha (2\mu_0)^{3/2}}{4} \Bigg[ \bigg( \frac{8}{3} - \frac{q^2}{3k_F^2} \bigg) \delta_{\alpha\beta} - \frac{q_\alpha q_\beta}{3k_F^2} \\
-& \frac{1}{3} \bigg( \frac{\Delta_Z}{2\mu_0} \bigg)^2 (\delta_{\alpha\beta} + \delta_{\alpha 3}) + \bigg( \frac{\Delta_Z}{2\mu_0} \bigg)^2 \bigg\{ \delta_{\alpha\beta} \frac{q^2}{30k_F^2} + \frac{q_\alpha q_\beta}{15k_F^2} \\
+& \delta_{\alpha\beta} (\delta_{\alpha 1} + \delta_{\alpha 2}) \frac{q^2 + q_z^2 + q_\alpha^2}{60k_F^2}  + (\delta_{\alpha 1} \delta_{\beta 2} + \delta_{\alpha 2} \delta_{\beta 1}) \frac{q_\alpha q_\beta}{60k_F^2} \bigg\} \Bigg].
\end{split}
\eeq
Here, terms of order $\mathcal{O}(\Delta_Z^4/(q^2/2m_b)^2)$ and higher are ignored.

The eigenmodes of the system are the roots of $\text{Det}[\hat{\mathcal{D}}^{-1}]=0$, where $\hat{\mathcal{D}}$ given in Eq.~\er{gr_t} accounts for phonon self-energy in Eq.~\er{Pi_q}. The resulting modes are:
\bse
\label{mode_q}
\beq
\begin{split}
\label{w1}
\Omega_1^2 = \omega_\text{TO}^2 & \bigg[ 1 - \frac{\alpha}{\alpha_0} \bigg\{ 1 - \frac{\Delta_Z^2}{8 (2\mu_0)^2} \bigg\} \bigg] \\
+ & \bigg[ c_T^2 + \frac{\alpha}{\alpha_0} \frac{\omega_\text{TO}^2}{8k_F^2} \bigg\{ 1 - \frac{3}{20} \frac{\Delta_Z^2}{(2\mu_0)^2} \bigg\} \bigg] \big( q_x^2 + q_y^2 \big) \\
+ & \bigg[ c_T^2 + \frac{\alpha}{\alpha_0} \frac{\omega_\text{TO}^2}{8k_F^2} \bigg\{ 1 - \frac{1}{5} \frac{\Delta_Z^2}{(2\mu_0)^2} \bigg\} \bigg] q_z^2, \\
\end{split}
\eeq
\beq
\label{w2}
\begin{split}
\Omega_2^2 = \omega_\text{TO}^2 & \bigg[ 1 - \frac{\alpha}{\alpha_0} \bigg\{ 1 - \frac{\Delta_Z^2}{8 (2\mu_0)^2} \bigg( 2 - \frac{q_z^2}{q^2} \bigg) \bigg\} \bigg] \\
+ & \bigg[ c_T^2 + \frac{\alpha}{\alpha_0} \frac{\omega_\text{TO}^2}{8k_F^2} \bigg\{ 1 - \frac{1}{10} \frac{\Delta_Z^2}{(2\mu_0)^2} \bigg\} \bigg] \big( q_x^2 + q_y^2 \big) \\
+ & \bigg[ c_T^2 + \frac{\alpha}{\alpha_0} \frac{\omega_\text{TO}^2}{8k_F^2} \bigg\{ 1 - \frac{1}{5} \frac{\Delta_Z^2}{(2\mu_0)^2} \bigg\} \bigg] q_z^2, \\
\end{split}
\eeq
\beq
\label{w3}
\begin{split}
\Omega_3^2 = \omega_\text{TO}^2 & \bigg[ 1 - \frac{\alpha}{\alpha_0} \bigg\{ 1 - \frac{\Delta_Z^2}{8 (2\mu_0)^2} \bigg( 1 + \frac{q_z^2}{q^2} \bigg) \bigg\} \bigg] \\
+ & \bigg[ c_L^2 + \frac{\alpha}{\alpha_0} \frac{\omega_\text{TO}^2}{4k_F^2} \bigg\{ 1 - \frac{1}{5} \frac{\Delta_Z^2}{(2\mu_0)^2} \bigg\} \bigg] \big( q_x^2 + q_y^2 \big) \\
+ & \bigg[ c_L^2 + \frac{\alpha}{\alpha_0} \frac{\omega_\text{TO}^2}{4k_F^2} \bigg\{ 1 - \frac{3}{20} \frac{\Delta_Z^2}{(2\mu_0)^2} \bigg\} \bigg] q_z^2. \\
\end{split}
\eeq
\ese
We emphasize again that the above result is obtained up to $\Delta_Z^2$ and $q^2$ order. Therefore, from Eq.~\er{mode_q} it appears that one can still go to $q\to0$ limit. This would be wrong because in the non-dispersive term of $\Omega_{1,2,3}$ (or massive terms), the higher order correction is of order $\mathcal{O}(\Delta_Z^4/(q^2/2m_b)^2)$. One cannot simply take $q\to0$ limit here given our initial assumption. Hence in the small $\Delta_Z$ regime, one must focus only on dispersive terms (terms of order $\sim q^2$). 

It is evident from Eq.~\er{mode_q} that for $\Delta_Z \ll 2\mu_0$ the dispersion of $\Omega_{1,2,3}$ never becomes negative, which indicates that the instability does not occur at finite $q$. One needs to go back to small $q$ regime to understand the instability at $q\to0$, which is discussed in Sec.~\ref{finite_q} of the main text.

\section{Scattering effects on the mode hybridization at $q=0$}
\label{app:damping}
In this section we will give technical details of the effect of scattering rate, characterized by $\Gamma$, on phonon self-energy at $q=0$ and $T=0$. The physical consequences of this on the mode hybridization is discussed in Sec.~\ref{disorder} of the main text. The electron Green's function in the spectral representation\cite{agd:1963, mahan:book} is written as
\beq
\label{spec_gr}
\frac{1}{i\omega_n - \xi_\bk - r \frac{\Delta_Z}{2}} = \int_{-\infty}^\infty \frac{d\Omega}{2\pi} \frac{A(\Omega, r)}{i\omega_n - \Omega}
\eeq
where
\beq
\label{spec}
A(\Omega, r) = \frac{2\Gamma}{(\Omega - \xi_\bk - r \Delta_Z/2)^2 + \Gamma^2}.
\eeq
is the spectral function. In this representation the phonon self-energy \er{TO_se} is written as
\beq
\label{dpza}
\begin{split}
\Pi_{ij}(\Omega_m) &= \frac{\lambda^2 \Omega_0^2}{8\pi^2} \sum_{s \bar{s}} \int_{-1}^1 d(\cos\theta) \int_0^{2\pi} \frac{d\phi}{2\pi} f_{ij}^{s \bar{s}} (\theta, \phi) \\
\times \int_0^\infty & \frac{dk}{2\pi} k^4 \int_{-\infty}^\infty \frac{d\Omega_1}{2\pi} \int_{-\infty}^\infty \frac{d\Omega_2}{2\pi} A(\Omega_1, \bar{s}) A(\Omega_2, s) \\
\times & T \sum_{\omega_n} \frac{1}{i\omega_n - \Omega_1} \frac{1}{i(\omega_n+\Omega_m) - \Omega_2},
\end{split}
\eeq
where $f_{ij}^{s \bar{s}}$ is the coherence factor given in Eq.~\er{coh_fac} of main text, and $A(\Omega, r)$ is the spectral function given in Eq.~\er{spec}. The frequency summation in Eq.~\ref{dpza} gives
\beq
T \sum_{\omega_n} \frac{1}{i\omega_n - \Omega_1} \frac{1}{i(\omega_n+\Omega_m) - \Omega_2} = \frac{n_F(\Omega_1) - n_F(\Omega_2)}{i\Omega_m - \Omega_2 + \Omega_1}.
\eeq
Now we take the imaginary part of $\Pi_{ij}$, which amounts to writing $1/[i\Omega_m - \Omega_2 + \Omega_1]$ as $-\pi \delta(\Omega - \Omega_2 + \Omega_1)$ according to the Sokhotski formula. Upon doing $\Omega_2$-integral, angle integration, and making a variable substitution from $k$ to $\xi_\bk$ according to the free electron dispersion relation, $k^2/2m_b - \mu = \xi_\bk$, we get at $T=0$:
\beq
\begin{split}
\text{Im} & \Pi_{ij}(\Omega) = -\frac{\alpha \Gamma^2}{\sqrt{2}} \sum_{s \bar{s}} h_{ij}^{s \bar{s}} \int_{-\mu}^\infty d\xi (\xi + \mu)^{3/2} \int_{-\Omega}^0 \frac{d\Omega_1}{2\pi} \times\\
\times & \frac{1}{(\Omega_1 - \xi - \bar{s} \Delta_Z/2)^2 + \Gamma^2} \frac{1}{(\Omega_1 + \Omega - \xi - s \Delta_Z/2)^2 + \Gamma^2},
\end{split}
\eeq
where $\xi \equiv \xi_\bk$ and
\beq
\begin{split}
h_{xx}^{s\bar{s}} = \int_{-1}^1 d(\cos\theta) \int_0^{2\pi} \frac{d\phi}{2\pi} f_{xx}^{s \bar{s}}(\theta, \phi) &= \frac{4}{3}, \\
h_{yy}^{s\bar{s}} = \int_{-1}^1 d(\cos\theta) \int_0^{2\pi} \frac{d\phi}{2\pi} f_{yy}^{s \bar{s}}(\theta, \phi) &= \frac{4}{3}, \\
h_{zz}^{s\bar{s}} = \int_{-1}^1 d(\cos\theta) \int_0^{2\pi} \frac{d\phi}{2\pi} f_{zz}^{s \bar{s}}(\theta, \phi) &= \frac{4}{3} (1-s\bar{s}), \\
h_{xy}^{s\bar{s}} = \int_{-1}^1 d(\cos\theta) \int_0^{2\pi} \frac{d\phi}{2\pi} f_{xy}^{s \bar{s}}(\theta, \phi) &= \frac{2i}{3} (s-\bar{s}).
\end{split}
\eeq
Since $\xi \ll \mu$, $(\xi + \mu)^{3/2} \approx \mu^{3/2}$ and the lower limit of the $\xi$-integral can be extended to $-\infty$.
After doing straightforward integral we obtain the imaginary part of $\Pi_{ij}$: 
\beq
\label{imdpz}
\begin{split}
\text{Im} & \Pi_{xx}(\Omega) = \text{Im}\Pi_{yy}(\Omega) = \frac{1}{2}\text{Im}\Pi_{zz}(\Omega) = \\
&= - \frac{\alpha(2\mu)^{3/2}}{12} \bigg[ \frac{4\Omega\Gamma}{(\Omega - \Delta_Z)^2 + 4\Gamma^2} + \frac{4\Omega\Gamma}{(\Omega + \Delta_Z)^2 + 4\Gamma^2} \bigg], \\
\text{Im} & \Pi_{xy}(\Omega) = -\text{Im}\Pi_{yx}(\Omega) = \\
&= - \frac{\alpha (2\mu)^{3/2}}{12} \bigg[ \frac{4\Gamma^2 - \Omega^2 + \Delta_Z^2}{(\Omega - \Delta_Z)^2 + 4\Gamma^2} - \frac{4\Gamma^2 - \Omega^2 + \Delta_Z^2}{(\Omega + \Delta_Z)^2 + 4\Gamma^2} \bigg].
\end{split}
\eeq
The real part can be calculated from the Kramers-Kronig relation, 
\beq
\label{KK}
\text{Re}\Pi_{ij}(\Omega) = \frac{1}{\pi} \mathcal{P} \int_{-\infty}^\infty d\Omega' \frac{\text{Im}\Pi_{ij}(\Omega')}{\Omega' - \Omega},
\eeq
where $\mathcal{P}$ denotes the Cauchy Principal value, and are written as
\beq
\label{redpz_main}
\begin{split}
\text{Re} & \Pi_{xx}(\Omega) = \text{Re}\Pi_{yy}(\Omega) = \frac{1}{2} \text{Re}\Pi_{zz}(\Omega) = \\
&- \frac{\alpha (2\mu)^{3/2}}{6} \bigg[ \frac{4\Gamma^2 - \Delta_Z (\Omega - \Delta_Z)}{(\Omega - \Delta_Z)^2 + 4\Gamma^2} + \frac{4\Gamma^2 + \Delta_Z (\Omega + \Delta_Z)}{(\Omega + \Delta_Z)^2 + 4\Gamma^2} \bigg] \\
\text{Re} & \Pi_{xy}(\Omega) = - \text{Re}\Pi_{xy}(\Omega) = \\
& \frac{\alpha (2\mu)^{3/2}}{12} \bigg[ \frac{4\Omega\Gamma}{(\Omega - \Delta_Z)^2 + 4\Gamma^2} - \frac{4\Omega\Gamma}{(\Omega + \Delta_Z)^2 + 4\Gamma^2} \bigg].
\end{split}
\eeq
The effect of resonance broadening, coming through self-energy, on hybridized modes is discussed in Sec.~\ref{disorder} of the main text.

\bibliography{referenceFile.bib}

\end{document}